\documentclass[12pt]{article}
\usepackage[latin1]{inputenc}
\usepackage{cite}
\usepackage{amsmath}
\usepackage{amsfonts}
\usepackage{amssymb}
\usepackage{graphicx}
\usepackage{geometry}
\usepackage{amssymb,epsfig}
\usepackage{hyperref}
\usepackage{slashed}
\usepackage{verbatim}
\usepackage{mathrsfs}

\graphicspath{{./}}
\allowdisplaybreaks

\makeatletter
\renewcommand\section{\@startsection {section}{1}{\z@}%
                                 {-3.5ex \@plus -1ex \@minus -.2ex}
                                   {2.3ex \@plus.2ex}%
                                   {\normalfont\large\bfseries}}
\renewcommand\subsection{\@startsection{subsection}{2}{\z@}%
                                   {-3.25ex\@plus -1ex \@minus -.2ex}%
                                     {1.5ex \@plus .2ex}%
                                     {\normalfont\bfseries}}
\renewcommand\subsubsection{\@startsection{subsubsection}{3}{\z@}%
                                   {-3.25ex\@plus -1ex \@minus -.2ex}%
                                     {1.5ex \@plus .2ex}%
                                     {\normalfont\itshape}}
\makeatother

\def\pplogo{\vbox{\kern-\headheight\kern -29pt
\halign{##&##\hfil\cr&{\ppnumber}\cr\rule{0pt}{2.5ex}&\ppdate\cr}}}
\makeatletter
\def\ps@firstpage{\ps@empty \def\@oddhead{\hss\pplogo}%
  \let\@evenhead\@oddhead 
}
\def\maketitle{\par
 \begingroup
 \def\thefootnote{\fnsymbol{footnote}}
 \def\@makefnmark{\hbox{$^{\@thefnmark}$\hss}}
 \if@twocolumn
 \twocolumn[\@maketitle]
 \else \newpage
 \global\@topnum\z@ \@maketitle \fi\thispagestyle{firstpage}\@thanks
 \endgroup
 \setcounter{footnote}{0}
 \let\maketitle\relax
 \let\@maketitle\relax
 \gdef\@thanks{}\gdef\@author{}\gdef\@title{}\let\thanks\relax}
\makeatother

\numberwithin{equation}{section}
\newcommand{\nn}{\nonumber}

\newcommand{\be}{\begin{equation}}
\newcommand{\bea}{\begin{eqnarray}}
\newcommand{\beq}{\begin{eqnarray}}
\newcommand{\ee}{\end{equation}}
\newcommand{\eea}{\end{eqnarray}}
\newcommand{\eeq}{\end{eqnarray}}

\newcommand{\lrf}[2]{\left(\frac{#1}{#2}\right)}
\newcommand{\nnmb}{\nonumber}

\newcommand{\gev}{\, {\rm GeV}}

\newcommand{\ccdot}{\!\cdot\!}
\newcommand{\lag}{\mathscr{L}}
\newcommand{\fh}{\varphi_1}
\newcommand{\ff}{\varphi_2}

\newcommand{\eg}{e.g.,\ }

\newcommand{\hc}{\mathrm{h.c.}}

\begin{document}
 
\setcounter{page}0
\def\ppnumber{\vbox{\baselineskip14pt
IPPP-15-23, DCPT-15-46}
}
\def\ppdate{\footnotesize{}} \date{}

\title{\bf 
Electroweak Baryogenesis from Exotic Electroweak Symmetry Breaking
}
\maketitle
\vspace{-2cm}

\begin{center}
\begin{large}
Nikita Blinov$^{(a,b)}$,
Jonathan Kozaczuk$^{(a)}$,\\
David E. Morrissey$^{(a)}$,
and Carlos Tamarit$^{(c)}$
\end{large}
\vspace{1cm}\\
  \begin{it}
(a) TRIUMF, 4004 Wesbrook Mall, Vancouver, BC V6T~2A3, Canada\vspace{0.3cm}\\
(b) Department of Physics and Astronomy, University of British Columbia,\\ 
Vancouver, BC V6T~1Z1, Canada\\
(c) Institute for Particle Physics Phenomenology, Department of Physics\\
Durham University, Durham DH1 3LE, United Kingdom
%
%
\vspace{0.5cm}\\
email: nblinov@triumf.ca, jkozaczuk@triumf.ca,\\ 
dmorri@triumf.ca, carlos.tamarit@durham.ac.uk\\
\vspace{0.2cm}
\end{it}
\today
\end{center}

\begin{abstract} \normalsize
\noindent 
We investigate scenarios in which electroweak baryogenesis can occur
during an exotic stage of electroweak symmetry breaking 
in the early Universe. This transition is driven by the expectation 
value of a new electroweak scalar instead of the standard Higgs field.  
A later, second transition then takes the system to the usual electroweak 
minimum, dominated by the Higgs, while preserving the baryon asymmetry
created in the first transition.  We discuss the general
requirements for such a two-stage electroweak transition to be suitable
for electroweak baryogenesis and present a toy model that illustrates
the necessary ingredients.  We then apply these results to construct
an explicit realization of this scenario within the inert
two Higgs doublet model.  Despite decoupling the Higgs from the
symmetry-breaking transition required for electroweak baryogenesis, 
we find that this picture generically
predicts new light states that are accessible experimentally.

\end{abstract}

\newpage

\section{Introduction}
 
 Electroweak baryogenesis~(EWBG) is an elegant mechanism for the generation of 
the baryon asymmetry during the electroweak phase transition~\cite{Trodden:1998ym,Cohen:1993nk,Morrissey:2012db}.  
If this phase transition is strongly first order, it proceeds
through the formation of bubbles of broken electroweak phase that
expand rapidly within the surrounding region of symmetric phase.  This departure
from thermodynamic equilibrium, together with $C$ (charge) and $CP$ (charge-parity) violation in particle
scattering at the bubble walls and $B+L$ violation by electroweak sphaleron
transitions, satisfy the necessary Sakharov conditions 
for baryogenesis~\cite{Sakharov:1967dj}.

  The standard realization of EWBG is driven by the Standard Model~(SM)
Higgs field (or a Higgs sector from which the SM-like Higgs emerges).
At high temperatures, the Higgs is stabilized at the symmetry-preserving
origin of its potential by thermal effects.  As the temperature
falls, the Higgs field develops a vacuum expectation value~(VEV) 
and breaks the manifest electroweak invariance.  EWBG requires
that this transition be strongly first-order, both to generate
bubble walls near which efficient baryon production can take place and to suppress
$B+L$ violation by sphaleron transitions
within the broken-phase bubbles.
The latter condition is more severe, and is usually quoted as
\beq
\frac{v_c}{T_c} \gtrsim 1 \ ,
\label{eq:nowash}
\eeq
where $v_c$ is the Higgs VEV at the critical temperature $T_c$
below which the first-order phase transition can occur.\footnote{
This condition is gauge dependent and has significant 
theoretical uncertainties~\cite{Patel:2011th} but is often
a good rule of thumb~\cite{Garny:2012cg}. 
We discuss gauge dependence and other uncertainties in more detail 
in Section~\ref{sec:gaugedep}}  Violation of $C$ and $CP$ in 
scattering with the Higgs bubble wall also emerges from interactions
with the changing Higgs field.

  This baryogenesis mechanism is particularly intriguing because the dynamics
takes place near the weak scale, suggesting that it can be probed
experimentally with current or upcoming data.  Indeed, EWBG has already
been tested and ruled out within the Standard Model~(SM), which contains
all the necessary ingredients.  The are two reasons for this: 
the electroweak phase transition~(EWPT) is not first order for the 
observed Higgs boson mass $m_h \simeq 125\,\gev$~\cite{Kajantie:1995kf,Kajantie:1996mn},
and the measured $CP$ violation from the Cabbibo-Kobayashi-Maskawa matrix 
is insufficient~\cite{Gavela:1993ts,Huet:1994jb,Gavela:1994dt}.  

  Electroweak baryogenesis is able to generate the observed baryon asymmetry 
in a number of extensions of the SM.  However, most of these realizations 
are becoming very strongly constrained as well.  To drive the EWPT to be
strongly first order, the new physics must couple significantly
to the SM Higgs, and this can lead to modifications to Higgs boson
production and decay rates.  For example, the EWPT can be made
strong in supersymmetric extensions of the SM through the effects
of a light scalar top~(stop) superpartner~\cite{Carena:1996wj,Delepine:1996vn}, 
by mixing with a new singlet
scalar~\cite{Davies:1996qn,Huber:2000mg,Menon:2004wv}, or even by arranging a
tree-level barrier between the Higgs vacuum and the origin.  
Comparing to data, the stop-driven scenario is mostly 
ruled-out by precision measurements of the Higgs production 
and decay rates at the 
LHC~\cite{Menon:2009mz,Cohen:2012zza,Curtin:2012aa,Carena:2012np} 
as well as by direct stop searches~\cite{Krizka:2012ah,Delgado:2012eu},
while Higgs measurements also limit the singlet-driven scenario to 
a small subset of the model parameter 
space~\cite{Huang:2014ifa, Kozaczuk:2014kva}. 
Coleman-Weinberg--like scenarios with tree-level barriers 
involve heavier exotic fields, 
but could be probed by precise measurements of 
the Higgs couplings~\cite{Espinosa:2008kw,Tamarit:2014dua}.
Similar conclusions are found in other extensions of the SM that
produce a strongly first-order EWPT~\cite{Chung:2012vg}.

  New sources of $CP$ violation are also needed for viable EWBG,
and these must connect to the SM Higgs so that they can be
enhanced by the advancing bubble walls.  This will typically
lead to new contributions to the permanent electric dipole moments (EDMs)
of leptons and baryons at the two-loop level~\cite{Pospelov:2005pr}.
In supersymmetric extensions, the main new source of $CP$ violation
in EWBG typically comes from the charginos and neutralinos,
and their contribution to EDMs is larger than the current
experimental limits except in a small resonantly 
enhanced window~\cite{Cirigliano:2009yd}.
Similar bounds are found in other theories~\cite{Shu:2013uua}.

  In this work, we investigate a non-standard realization of
electroweak baryogenesis involving a two-stage phase transition
as a means to sidestep some of these constraints, making use of the
mechanism first introduced in Ref.~\cite{Patel:2012pi}.  
The key idea is to have an initial stage of electroweak symmetry breaking 
induced by the VEV of an exotic $SU(2)_L$-charged scalar rather 
than the standard Higgs.
At some later time, the system will evolve from the exotic 
vacuum to the standard Higgs vacuum that we occupy today.
Baryon production via EWBG could then occur in the first, exotic field
transition.  As long as the transition to the standard Higgs vacuum
maintains enough electroweak symmetry breaking 
(or if it is strongly first-order),
the baryons created in the initial phase transition will be preserved.
Additional new physics (beyond the exotic scalar) needed to drive the strong
initial transition or to generate $CP$ violation can couple primarily to the
exotic scalar rather than the SM-like Higgs field.  This suggests that
the bounds on EWBG can be relaxed in this scenario.
We will see that this is only partially true.

  Multi-stage phase transitions have been considered before in the context
of EWBG.  In scenarios with a singlet scalar that helps to drive a strongly
first-order Higgs phase transition, the singlet is often found to develop
a VEV before the 
Higgs field~\cite{Ahriche:2007jp,Profumo:2007wc, Chung:2012vg, Huang:2014ifa,Kozaczuk:2014kva,Curtin:2014jma}.  
Electroweak symmetry breaking by an exotic
doublet was suggested in Ref.~\cite{Land:1992sm} 
with the goal of inducing EWBG in a strongly first-order phase transition 
from the exotic vacuum to the Higgs vacuum.  
This does not work because of a strong suppression of sphalerons
in the initial exotic broken phase~\cite{Hammerschmitt:1994fn}.

Two-stage electroweak symmetry breaking with EWBG taking place in
the first step was proposed in Ref.~\cite{Patel:2012pi}. 
Here, we aim to investigate and extend this mechanism further. 
Relative to Ref.~\cite{Patel:2012pi}, which was based on a triplet extension 
of the SM, we elucidate several general criteria for such two-stage 
electroweak phase transitions, and we exhibit a new realization 
suitable for EWBG within an inert two-Higgs-doublet model.
We also discuss the resulting particle phenomenology of this class of theories.

The outline of this paper is as follows. 
In Section~\ref{sec:requirements}, we argue that two-stage electroweak 
phase transitions are difficult to achieve with one field alone, 
generically requiring additional fields transforming non-trivially 
under $SU(2)_L$. To illustrate what this entails, we present 
a toy model that realizes a two-stage phase transition in a simple 
and general way.  In Section~\ref{sec:2HDM}, we apply these results 
to a realistic inert two-Higgs-doublet model that exhibits a two-step
electroweak symmetry--breaking transition in which the first stage 
is suitable for EWBG.
The experimental constraints on this scenario are studied 
in Section~\ref{sec:constraints}, where we show that an additional 
contribution to the masses of the exotic scalars is needed, and we
exhibit a simple extension of the inert doublet model in which this can occur.
Finally, Section~\ref{sec:conc} is reserved for our conclusions.
Some technical details about our treatment of the effective potential 
and its thermal corrections are collected in a pair of appendices.

\section{\label{sec:requirements}Requirements for a Viable Two-Step Transition}

We are interested in scenarios in which electroweak symmetry
breaking occurs in two stages in the early Universe, 
with the first transition proceeding towards an exotic vacuum
and the second transition bringing the system (close) to the standard
electroweak minimum. In this section we outline the general conditions under 
which such a two-stage transition can occur in a realistic way,
list the additional requirements for successful EWBG, 
and present a simplified toy model that illustrates what is needed.

\subsection{General Conditions with One Electroweakly Charged Scalar}

   A minimal scenario for a two-stage transition would involve  
a new exotic vacuum in the potential of a single Higgs scalar doublet.  
At zero temperature the exotic vacuum should lie above 
the standard Higgs vacuum or disappear entirely, while at high temperature 
the SM vacuum should be lifted to enable a direct transition 
from the symmetry-preserving origin to the exotic vacuum.
This reversal in the ordering of the free-energies of the vacua implies 
that the effective potential in the region of the standard vacuum needs 
to receive stronger thermal corrections than the potential 
in the region of the exotic minimum.  
We do not expect this to occur if the false vacuum has a larger VEV 
than the SM minimum, since thermal corrections to the potential increase 
for growing values of the fields, until they become roughly constant due to
Boltzmann suppression.\footnote{The growth of thermal corrections for
small enough values of the fields can be seen by noting that the
leading thermal corrections to the effective potential are given 
by positive temperature-dependent mass terms, quadratic in the fields.}
If, on the other hand, the new vacuum lies closer to the origin 
-- which is difficult to realize in a perturbative setting -- 
it would be challenging to obtain $v_c/T_c\gtrsim 1$ with both 
a smaller VEV and a higher critical temperature than 
the ones corresponding to the standard Higgs vacuum. For these reasons
we will not consider this possibility further.

\subsection{\label{subsec:general2scalars}General Conditions with Two Electroweakly Charged Scalars}

  For the remainder of this work we will focus on theories with 
two electroweakly-charged scalars $\Phi$ and $H$, with $\Phi$ mostly inert and 
$H$ associated with the standard electroweak vacuum, 
gauge boson and fermion masses, and the Higgs boson.  
At zero temperature, $T=0$, 
electroweak symmetry breaking should therefore be dominated by the 
VEV of $H$.  A VEV for $\Phi$ is also possible, but it must be much smaller 
than the VEV of $H$ to avoid large corrections to electroweak 
observables or Higgs boson production 
rates~\cite{Enberg:2013ara,Enberg:2013jba}.  
We will focus on the case where $\langle\Phi\rangle$ 
is negligible today.  Thus, any non-standard vacua present at $T=0$
should be shallower than the Higgs vacuum, or separated by a large
barrier.

  At very high temperatures, thermal corrections are expected
to drive the ground state of the system to the origin of
the $(H,\Phi)$ field space.  To realize the two-stage scenario
we are interested in, the exotic $\Phi$ field must
develop a VEV first.  This can occur readily if the $H$ field 
has larger couplings than $\Phi$ to light matter fields in 
the cosmological plasma, since these will provide 
a stronger thermal stabilization of the $H$ direction.  
If EWBG is to occur in the first transition along the $\Phi$ direction, 
it must also be strongly first order to suppress non-perturbative
baryon washout processes such as sphaleron 
transitions~\cite{Arnold:1987mh,Quiros:1999jp}
or electroweak monopoles~\cite{Patel:2012pi}.

\begin{figure}[ttt]\centering
  \includegraphics[scale=.28]{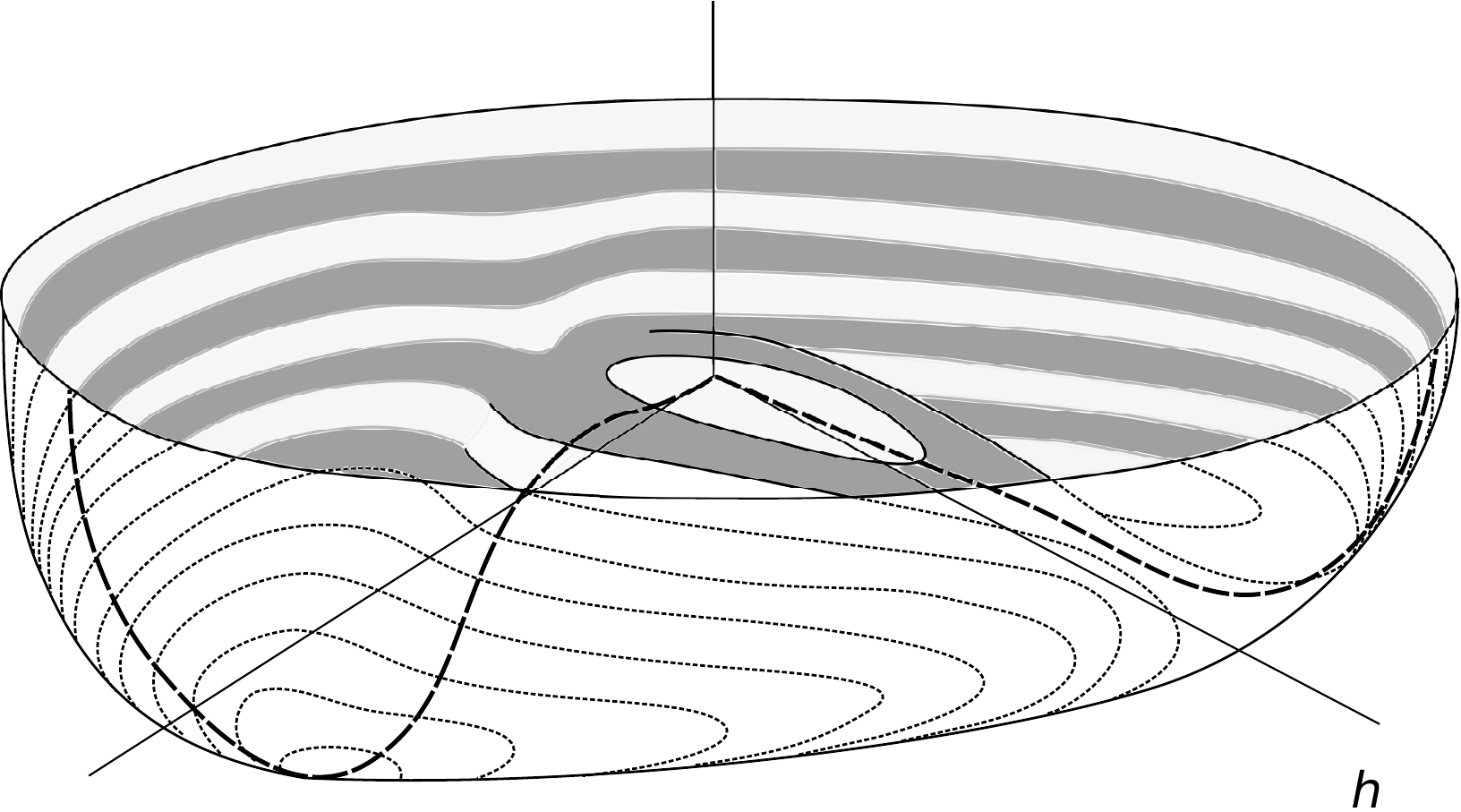}
  \hspace{0.2cm}
  \includegraphics[scale=.32]{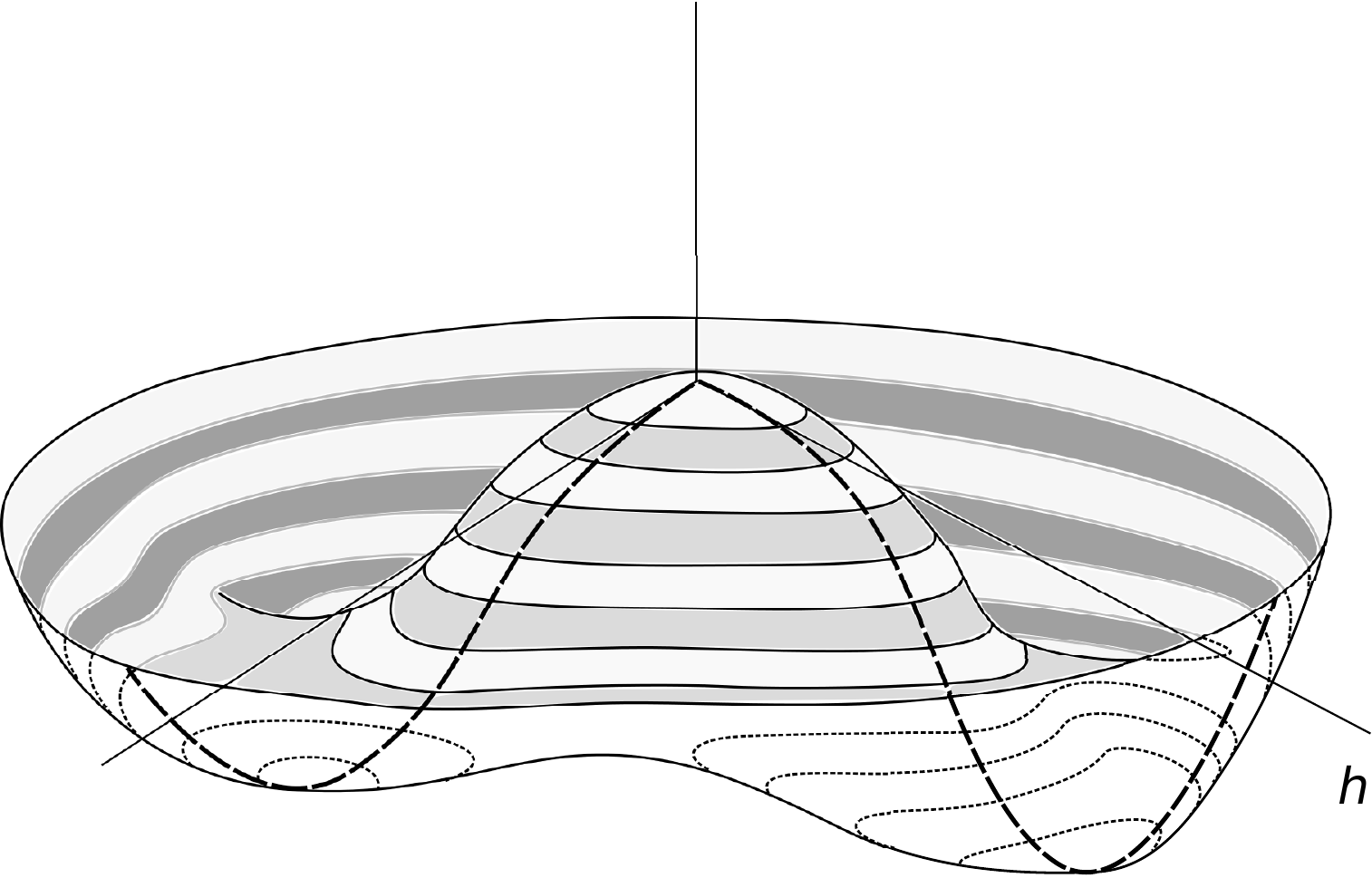}
  \hspace{0.2cm}
  \includegraphics[scale=.32]{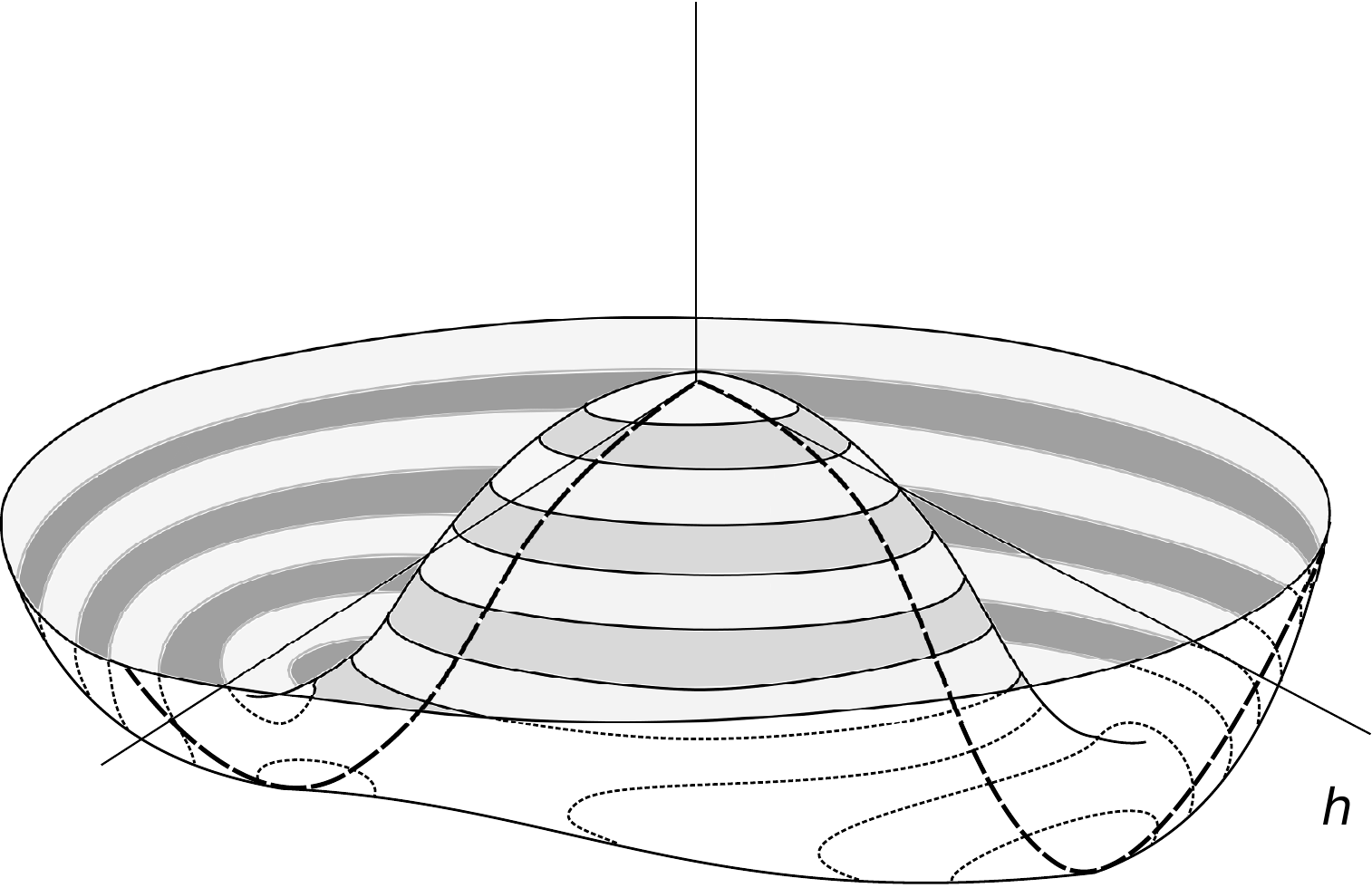}  

\caption{\label{fig:multifieldpot}Scenarios with vacua aligned along different
field directions. Left: High temperature potential with no vacuum in the Higgs
direction, enabling a first-order transition to the exotic vacuum. Middle and
Right: Possible low or zero temperature potentials, for which first- (middle)
or second-order (right) transitions towards the Higgs vacuum are possible.
}
\end{figure}

  As the temperature falls further, the system should evolve 
from the exotic vacuum to something close to the standard 
electroweak minimum dominated by the VEV of $H$,
as shown in Fig.~\ref{fig:multifieldpot}.
This later transition can be first- or second-order.  
However, if it is to preserve the baryon number
generated by EWBG in the initial transition, it must satisfy some
additional conditions.  When the second transition is first-order,
corresponding to the middle panel of Fig.~\ref{fig:multifieldpot},
it must complete efficiently enough that it does not inject a
large amount of entropy that would overly dilute the baryon
asymmetry~\cite{Patel:2012pi}.  If the second transition is second-order,
as in the right panel of Fig.~\ref{fig:multifieldpot},
the total amount of electroweak symmetry breaking must be large 
enough to suppress baryon washout by sphaleron 
transitions~\cite{Arnold:1987mh,Quiros:1999jp}.  
In the case of two electroweak doublets, this translates into a condition 
of the form~\cite{Ahriche:2014jna}
\beq
\frac{\sqrt{\langle H\rangle^2+\langle\Phi\rangle^2}}{T} ~\gtrsim~ 1/\sqrt{2} \ ,
\eeq
where it should be noted that there is a significant uncertainty
due to the gauge dependence of the effective potential and higher-order
parts of the sphaleron transition rate~\cite{Patel:2011th}.
A similar condition applies when the exotic multiplet occupies a higher-dimensional representation of $SU(2)_L$~\cite{Ahriche:2014jna}.

\subsection{A Two-Stage Toy Model\label{sec:toy}}

  The general conditions for a two-stage electroweak transition
described above can be understood simply in a toy model containing 
a pair of complex scalar fields.  While this model does not contain 
any gauge degrees of freedom, and the first transition
is usually not first-order, it will serve as a useful reference
for studying realistic theories of electroweak symmetry breaking
and EWBG below.  

  The model consists of two complex scalar fields $\Phi_1$ and $\Phi_2$
and $N$ copies of a ``top'' fermion $T$ with Lagrangian
\beq
-\lag &=& -\lag_{\rm kinetic}
\label{eq:ltoy}\\
&&-\mu_1^2|\Phi_1|^2 - \mu_2^2|\Phi_2|^2 
+ \frac{\lambda_1}{2}|\Phi_1|^4 + \frac{\lambda_2}{2}|\Phi_2|^4
+ {\lambda_3}|\Phi_1|^2|\Phi_2|^2 \nnmb\\
&&
+ \left(y_T\Phi_1\overline{T}_LT_R + \hc\right) \ ,\nnmb
\eeq
with $\mu_i^2 > 0$ and $\lambda_i>0$.
The potential of this theory has two independent $U(1)$ global symmetries under which
each of the complex scalars transforms independently.  
The negative quadratic interactions will induce VEVs 
for one or both fields.  Choosing a vacuum
where both VEVs are real, we expand the fields as
\beq
\Phi_1 &=& (\fh+r_1+iA_1)/\sqrt{2} \\
\Phi_2 &=&(\ff+r_2+iA_2)/\sqrt{2} 
\eeq
where $\varphi_{1,2}$ are the (canonically-normalized) background 
scalar fields entering the effective potential.
The tree-level effective potential then becomes
\beq
V_0(\fh,\ff) &=& -\frac{1}{2}\mu_1^2\fh^2-\frac{1}{2}\mu_2^2\ff^2 
+ \frac{\lambda_1}{8}\fh^4+\frac{\lambda_2}{8}\ff^4
+\frac{\lambda_3}{4}\fh^2\ff^2 \ ,
\label{eq:vtoy}\\
&=& -\frac{1}{2}(\mu_1^2\fh^2+\mu_2^2\ff^2) 
+ \frac{\lambda_3}{8\mu_1^2\mu_2^2}\left(\mu_1^2\fh^2+\mu_2^2\ff^2\right)^2
+ \frac{\Delta\lambda_1}{8}\fh^4+\frac{\Delta\lambda_2}{8}\ff^4 \ , \nnmb
\eeq
where the second line is a suggestive rewriting of the potential in terms of
\beq \label{eq:dlambda}
\Delta\lambda_1 = \lambda_1 -\frac{\mu_1^2}{\mu_2^2}\lambda_3  \ ,~~~~~
\Delta\lambda_2 = \lambda_2 - \frac{\mu_2^2}{\mu_1^2}\lambda_3  \ .
\eeq

For any values of $\Delta\lambda_1$ and $\Delta\lambda_2$, 
the potential always has local extrema 
at $(v_1,0)$ and $(0,v_2)$ 
with $v_i^2 = 2\mu_i^2/\lambda_i$ and $V(v_i) = -\mu_i^4/2\lambda_i$.  
In the special case of $\Delta\lambda_1=\Delta\lambda_2 = 0$, 
these local extrema both coincide with an ellipse of minima in
$\fh\!-\ff$ plane defined by 
$(\mu_1^2\fh^2+\mu_2^2\ff^2) = 2\mu_1^2\mu_2^2/\lambda_3$.
Turning on small values of $\Delta\lambda_1$ and $\Delta\lambda_2$, the ellipse
is deformed and the potential develops discrete minima.  
These minima may coincide with the \emph{exclusive} extrema 
at $(v_1,0)$ and $(0,v_2)$, or they may lie in a valley connecting them.  
Examining the local stability conditions for the exclusive extrema, 
$(v_1,0)$ is stable for $\Delta\lambda_1 < 0$ and $(0,v_2)$ 
is stable for $\Delta\lambda_2 < 0$.
This implies three distinct cases (up to the exchange $1\leftrightarrow 2$):
\vspace{-0.2cm}
\begin{enumerate}
\item $\mathbf{\Delta\lambda_1 <0,~\Delta\lambda_2 > 0}$\\
$(v_1,0)$ is a minimum and $(0,v_2)$ is unstable, and they are
connected by a smoothly decreasing valley.
\item $\mathbf{\Delta\lambda_1 <0,~\Delta\lambda_2 < 0}$\\
$(v_1,0)$ and $(0,v_2)$ are both local minima, and there exists
a higher saddle-point barrier within the valley connecting them.
\item $\mathbf{\Delta\lambda_1 >0,~\Delta\lambda_2 > 0}$\\
$(v_1,0)$ and $(0,v_2)$ are both saddle points, and there exists
a deeper minimum within the valley connecting them.
\end{enumerate}
\vspace{-0.5cm}
We will be primarily interested in Cases 1 and 2; Case 3 is undesirable
because both fields would obtain non-negligible VEVs.

At finite temperature $T$, the effective scalar potential will receive
thermal corrections.  The leading effect at large $T$ is to modify the scalar
squared-mass parameters in Eq.~\eqref{eq:vtoy} according to
\beq
\mu_i^2 ~\to~ m_i^2 = \mu_i^2 - a_iT^2 
\eeq
with 
\beq \label{eq:a}
a_1 = \frac{\lambda_1}{6} + \frac{\lambda_3}{12} + N\frac{y_T^2}{12} \ ,~~~~~
a_2 = \frac{\lambda_2}{6} + \frac{\lambda_3}{12}  \ .
\eeq
At very high $T$, both $m_1^2$ and $m_2^2$ are negative
and the only minimum of the potential lies at the origin.
As the temperature falls, one of the $m_i^2$ will become positive
and a VEV will develop along the corresponding field direction.
For appropriate values of the parameters, this theory may also
undergo a second, later transition in which the other field develops
a VEV.

  To illustrate a two-stage transition, let us take $\Delta\lambda_1 < 0$
and $\Delta\lambda_2 > 0$ (at $T=0$) as in Case~1 above, 
so that the minimum at $T=0$ lies at $(v_1,0)$.
The $\varphi_1$ and $\varphi_2$ VEVs are then analogous to the SM-like Higgs 
and the exotic scalar fields, respectively.  For the first transition
to occur exclusively in the $\ff$ direction, $m_2^2$ should become positive
before $m_1^2$, which requires
\beq \label{eq:destabilize}
\mu_2^2/a_2 > \mu_1^2/a_1 \ .
\eeq
The inequality can be satisfied if the coupling $y_T$ in Eq.~\eqref{eq:ltoy} 
is large enough or $\lambda_2$ is small enough.  
As the temperature falls further, the minimum in the $\ff$ direction 
will disappear and the field will end up in the $T=0$ minimum at $(v_1,0)$. 
This can happen smoothly in a second-order transition, 
or via tunneling in a first-order transition if a barrier 
is generated by thermal effects. 

  A two-step transition can occur in a similar way in Case~2,
with $\Delta\lambda_1 < 0$ and $\Delta\lambda_2 < 0$.  However,
a barrier between the $\varphi_1$ and $\varphi_2$ vacua is present 
at tree-level in this case and persists at $T=0$.  
The second transition is now first order, and there is a danger
that it never completes or that it will lead to strong supercooling
followed by a substantial injection of entropy.  Both outcomes are problematic
for electroweak baryogenesis, and we will study them further in
the more realistic theories to be considered below.

\begin{figure}[ttt]
\centering
  \includegraphics[width=8cm]{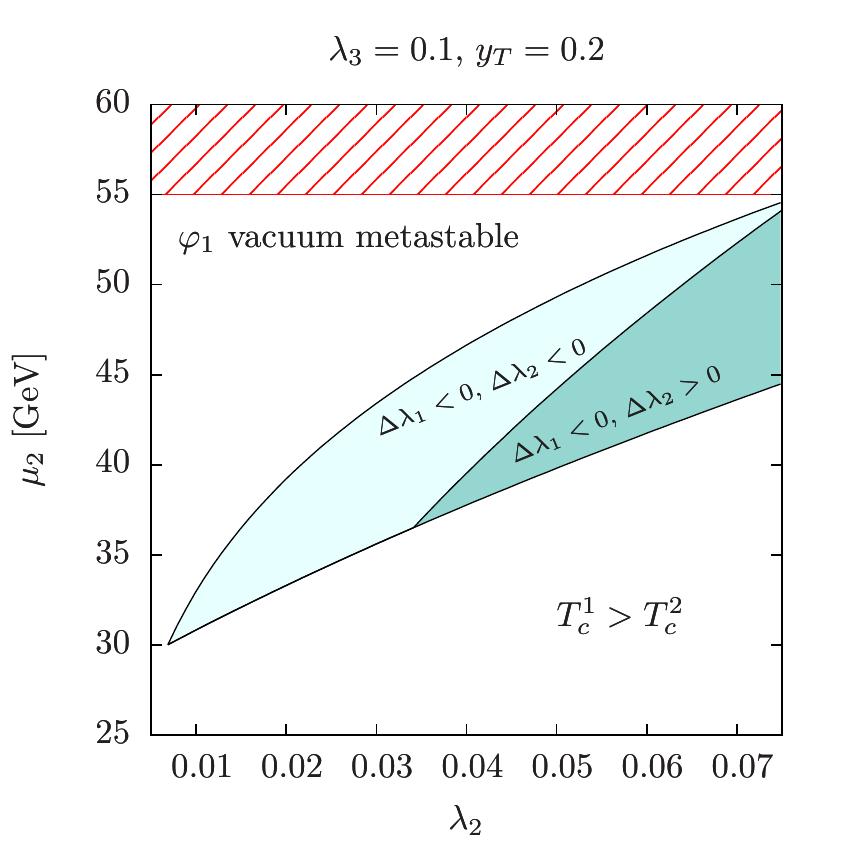}
\caption{\label{fig:toy} 
Parameter regions in $\lambda_2$--$\mu_2$ plane of the toy model where 
a two-stage phase transition is likely to occur, first in the ``exotic''
$\varphi_2$ direction and then proceeding to the ``standard'' 
$\varphi_1$ direction.  The regions correspond to the conditions 
of Eqs.~(\ref{eq:toy1}, \ref{eq:toy2}, \ref{eq:toy3}, \ref{eq:toy4}),
with $\lambda_3=0.1$, $y_T=0.2$ with $N=3$, 
and $\lambda_1$ and $\mu_1^2$ set as described in the text.
All four conditions are satisfied in the darker shaded region.}
\end{figure}

  Many features of this simple toy model will be applicable to
extended Higgs sectors giving rise to two-stage transitions where the
VEVs are aligned along one of the field directions at each of the transitions.
In particular, we can derive four conditions on the parameter space of
the toy model that will serve as a useful guide for realistic theories
when we identify $\Phi_1$ with the standard Higgs doublet and $\Phi_2$
with an exotic electroweak scalar multiplet.
Fixing $\lambda_1$, $\lambda_3$, and $\mu_1^2$, these conditions can be expressed
as inequalities in the $\lambda_2$--$\mu_2$ plane:
\beq
\mu_2^2 &<& \lrf{\lambda_3}{\lambda_1}\mu_1^2~~~~~\,(\Delta\lambda_1 < 0) 
\label{eq:toy1}\\
\mu_2^2 &<& \sqrt{\frac{\lambda_2}{\lambda_1}}\mu_1^2~~~~~~~(V(v_1)<V(v_2))
\label{eq:toy2}\\
\mu_2^2 &>& \lrf{a_2}{a_1}\mu_1^2~~~~~~(T_c^{2}> T_c^{1})
\label{eq:toy3}\\
\mu_2^2 &<& \lrf{\lambda_2}{\lambda_3}\mu_1^2\;~~~~~(\Delta\lambda_2 > 0) \ .
\label{eq:toy4}
\eeq
The first condition, Eq.~\eqref{eq:toy1}, is a restatement of 
$\Delta\lambda_1 < 0$, which is necessary for the ``standard'' 
electroweak minimum $(v_1,0)$ to be locally stable.  
The second condition, Eq.~\eqref{eq:toy2}, ensures that this minimum 
is a global minimum at $T=0$, deeper than the exotic minimum at $(0,v_2)$.
The third condition, Eq.~\eqref{eq:toy3}, is equivalent to 
Eq.~\eqref{eq:destabilize} and implies that the exotic $\varphi_2$ direction
is likely to be destabilized before the $\varphi_1$ direction.
The fourth condition, Eq.~\eqref{eq:toy4}, corresponds to $\Delta\lambda_2 > 0$ 
and the absence of a barrier between $(v_1,0)$ and $(0,v_2)$ 
at zero temperature.  

  In Fig.~\ref{fig:toy}, we illustrate the region satisfying all these
conditions.  We fix $\lambda_3=0.1$, $y_T =0.2$ with $N=3$, and
set $\lambda_1$ and $\mu_1^2$ such that $v_1 = 246\,\gev$ and
$m_{r_1} = 125\,\gev$ (where $r_1$ is the real 
scalar fluctuation around $(v_1,0)$).  The upper hatched region is excluded
because it lacks a stable electroweak minimum.  In the shaded regions,
the conditions of Eqs.~(\ref{eq:toy1}, \ref{eq:toy2}, \ref{eq:toy3}) are satisfied,
with the requirement of Eq.~\eqref{eq:toy4} also realized in the darker
shaded region. A bigger portal coupling $\lambda_3$ would lead to a larger tree-level 
barrier and therefore decrease the area of this region via Eq.~\eqref{eq:toy4}.
Increasing the value of $y_T$ or $N$ enhances thermal 
corrections in the $\phi_1$ direction, making it easier to satisfy $T_c^1 < T_c^2$ (see Eq.~\eqref{eq:toy3}), which 
enlarges the wedge region. We will see below that when these considerations
are applied to realistic theories of electroweak symmetry breaking,
the parameter regions supporting a two-stage transition amenable to EWBG
mostly correspond to the darker shaded region in this figure, 
where all four conditions are satisfied.

\section{Two Stages with Two Higgs Doublets\label{sec:2HDM}}

Armed with the general considerations of the previous section, 
we turn next to realistic Higgs sectors, focusing primarily 
on a restricted inert two-Higgs-doublet model.  
We show that a strongly first-order electroweak-symmetry-breaking phase
transition can occur first in the direction of the inert doublet,
followed by an efficient second transition to the standard Higgs
vacuum.  We also discuss certain technical aspects of the calculation,
and comment on a related extension with a new electroweak triplet scalar. 
 
 \subsection{Methodology}\label{sec:method}

  Most of our analysis will focus on the inert doublet model, 
consisting of the SM Higgs doublet $H$ together with 
an inert doublet $\Phi$ with no direct couplings to the SM fermions.  
The tree-level scalar potential is
\beq
\nonumber V(H,\Phi)&=&-m^2_H H^\dagger H-m^2_\Phi \Phi^\dagger\Phi+\frac{\lambda}{2}(H^\dagger H)^2+\frac{\lambda_\Phi}{2}(\Phi^\dagger \Phi)^2
\label{eq:2HDMsterile}
\\
&&~+\lambda_{\Phi H}(H^\dagger H)(\Phi^\dagger \Phi)
+\tilde\lambda_{\Phi H}(\Phi^\dagger H)(H^\dagger \Phi) 
\nnmb\\
&&~+ \frac{\lambda_5}{2}\left[(\Phi^{\dagger}H)^2 + \hc\right] \ .\nnmb
\eeq
This is the most general potential allowed when $\Phi$ is odd under
a $\mathbb{Z}_2$ symmetry. This parity also forbids renormalizable 
couplings of $\Phi$ to SM fermions.
A larger global $U(1)$ symmetry acting on $\Phi$
is present in the limit $\lambda_5\to 0$~\cite{Hambye:2007vf}.  
If both $\lambda_5 = 0 = \tilde{\lambda}_{\Phi H}$,
the global symmetry is enhanced further to an $SU(2)$ 
acting on $\Phi$~\cite{Hambye:2007vf}.\footnote{This global $SU(2)$
applies to the scalar potential, but is broken explicitly in the full
Lagrangian by the gauging of $SU(2)_L$.  Even so, the global
$SU(2)$ can still be a useful approximate symmetry.}

  Motivated by the SM-like Higgs boson observed at 
the LHC~\cite{Aad:2014aba,Khachatryan:2014jba} 
and the consistency of the SM with precision electroweak tests, 
we will focus exclusively on parameters where only $H$ has a VEV today:
$\langle H\rangle = v/\sqrt{2} = (246/\sqrt{2})\,\gev$
and $\langle \Phi\rangle = 0$.  Negative quadratic
mass parameters are assumed (corresponding to $m_H^2,\,m_{\Phi}^2 > 0$),
as well as positive values of $\lambda$, $\lambda_{\Phi}$,
and $\lambda_{\Phi H}$.  For simplicity, we also take $\lambda_5\to 0$,
and $\tilde{\lambda}_{\Phi H} < 0$ with 
$|\tilde\lambda_{\Phi H}| \ll \lambda_{\Phi H}$. 
We will see that these choices are close to optimal for generating a two-stage
electroweak phase transition, while a small but negative value
of $\tilde{\lambda}_{\Phi H}$ also forces the charged components
of the exotic doublet to be slightly heavier than the neutral ones.  

The thermal evolution of the theory is tracked by computing
the one-loop thermal effective potential $V^1_{\rm eff}(h,\phi,T)$
in the Landau gauge ($\xi = 0$), where $h$ and $\phi$ are 
the canonically-normalized background 
field values corresponding to the real neutral components 
of $H\to h/\sqrt{2}$ and $\Phi\to \phi/\sqrt{2}$.  
We further improve the effective potential using the renormalization group
with two-loop beta functions and anomalous dimensions when determining the
couplings in terms of experimental measurements at scales $m_Z$ and $m_t$,
with the relevant (one-loop) expressions for these collected in
Appendix~\ref{app:RG}.
The top Yukawa coupling is determined from the pole mass 
including SM strong and weak threshold corrections 
of order $\alpha_s^3, \alpha_w$. 
The Higgs quartic and mass parameters are fixed by requiring the correct
Higgs VEV and physical mass, which we take to be $v=246\,\gev$ 
and $m_h = 125\gev$~\cite{Aad:2015zhl}. 
The latter is matched to a zero of the neutral Higgs 1PI two-point 
function at $p^2=m_h^2$, the 1PI function being equal to
the second derivative of the effective potential with respect to $h$ (yielding
the zero momentum contribution) plus a finite momentum correction obtained from
the general formulae of Ref.~\cite{Martin:2003it}. 

  The negative quadratic coefficients of $H$ and $\Phi$ generate
imaginary parts in the effective potential at low temperatures and small field values.
To minimize these effects, we resum the one-loop zero-temperature corrections 
for the Goldstone self-energies~\cite{Martin:2014bca,Elias-Miro:2014pca} 
as well as the dominant finite temperature corrections for all bosonic 
self-energies~\cite{Parwani:1991gq,Arnold:1992rz}. 
The thermal \emph{daisy} resummation is known to improve convergence 
of the thermal corrections for small values of the effective 
bosonic masses; details of the temperature corrections 
to the bosonic masses are given in Appendix~\ref{app:T}.

  Second-order phase transitions are straightforward to analyze using
only the thermal effective potential, but first-order transitions require
more care.  Such transitions can occur when two local minima, $(h,\phi) = (v_1,w_1)$ and $(v_2,w_2)$, 
are present at the same temperature.
In this case, the critical temperature $T_c$ is defined by the relation
\beq
\label{eq:Tc}
V^1_{\rm eff}(v_1,w_1;T_c)= V^1_{\rm eff}(v_2,w_2;T_c) \ ,
\eeq
where $V^1_{\rm eff}$ includes the resummations described above.
Suppose the system starts in the $(v_2,w_2)$ vacuum just above $T_c$, 
and the $(v_1,w_1)$ minimum becomes deeper than the first as $T$ falls below
$T_c$.  In this case, the system may transition to the second vacuum
at some lower temperature $T_n$ by thermal fluctuations or quantum
tunneling through the nucleation of bubbles of the lower-energy phase.  

  We analyze such tunneling transitions using the 
\texttt{CosmoTransitions} package~\cite{Wainwright:2011kj},
modified to ensure that the thermal corrections are evaluated appropriately
for negative values of their arguments.
When more than one minimum is present, the package is also used to
compute the multifield tunneling rates and nucleation temperatures.
The nucleation temperature of a first-order transition is the temperature 
below which the bubble nucleation rate overcomes the expansion rate 
of the Universe, allowing the transition to complete.
When thermal tunneling dominates the formation 
of bubbles, this occurs when 
$S_3/T_n \lesssim 140$~\cite{Anderson:1991zb}, where $S_3$ 
is the three-dimensional Euclidean action of the instanton corresponding 
to a thermal transition between the relevant vacua. In all cases, 
the relevant temperatures are large enough such that thermal
transitions are more efficient than $O(4)$-symmetric
quantum tunneling.

  Let us also mention that two-stage electroweak phase transitions have
been studied in the inert two-Higgs doublet model 
in Ref.~\cite{Ginzburg:2010wa} and more generally in Ref.~\cite{Jarvinen:2009wr}.
It was also noted in Ref.~\cite{Gil:2012ya} that non-inert vacua 
may appear due to large quantum corrections. 
The inert doublet model has also been applied to generate a strongly first-order
Higgs-driven one-step electroweak phase transition suitable for EWBG
in Refs.~\cite{Chowdhury:2011ga,Borah:2012pu,Gil:2012ya,Cline:2013bln,AbdusSalam:2013eya}.  As far as we can tell, EWBG in a two-step transition 
has not been studied previously in this context.

\subsection{Parameter Scans\label{subsec:sterile}}
 
  Following the methods described above, we have analyzed the
inert doublet model of Eq.~\eqref{eq:2HDMsterile} by scanning over the 
model parameters and searching for acceptable two-stage phase
transitions.  For each parameter point, we start at a very high
temperature and track the evolution of the potential down
to determine the order and structure of the cosmological
phase transitions. The free parameters in our scans are 
$m_{\Phi}^2$, $\lambda_{\Phi}$, $\lambda_{\Phi H}$, and $\tilde\lambda_{\Phi H}$, 
with the other parameters fixed in terms of $m_h$ and $v$ 
to guarantee a (local) minimum at $T=0$ with $\phi = w =0$,
$h = v=246\,\gev$, and $m_h = 125\,\gev$.  
We focus on $\lambda_{\Phi} \in [0.01,0.055]$, 
$\sqrt{m_{\Phi}^2} \in [38,60]\,\gev$,
for fixed values of $\lambda_{\Phi H} \in \{0.1,\,0.2\}$ and
$\tilde\lambda_{\Phi H} = -0.001$.

\begin{figure}[ttt]
\centering
  \includegraphics[width=0.47\textwidth]{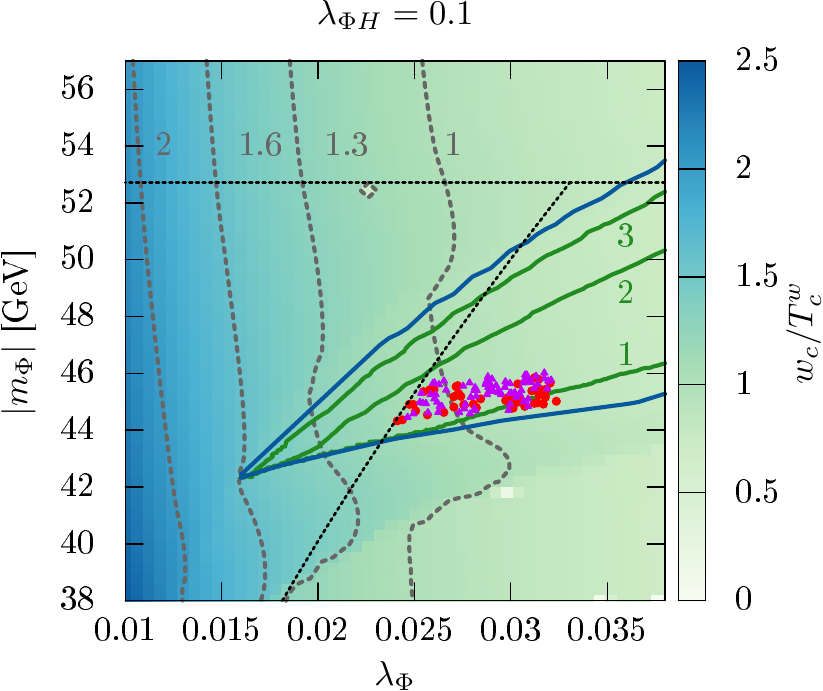}
  \includegraphics[width=0.47\textwidth]{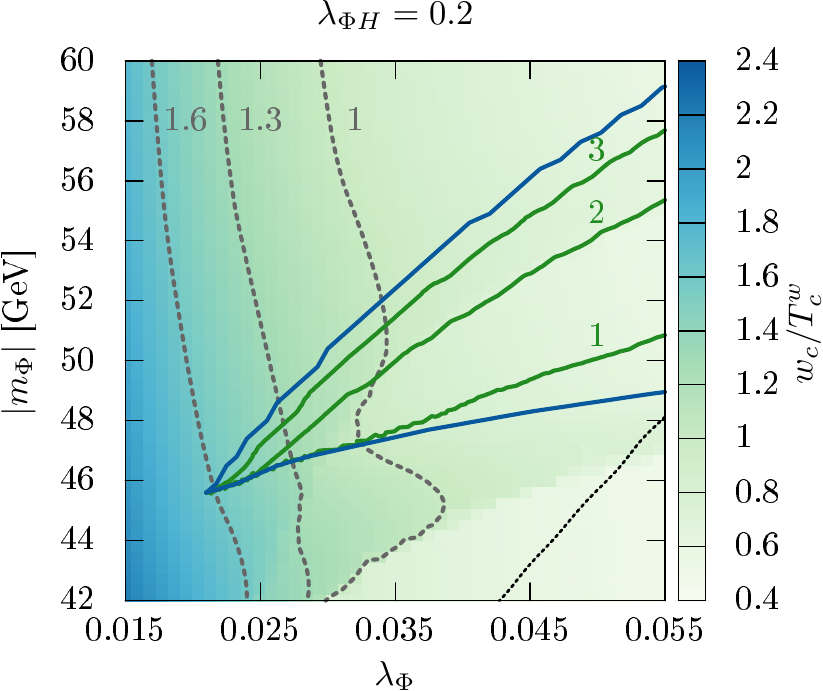}
\caption{\label{fig:sterile}
Parameter points leading to a two-stage electroweak phase transition amenable
to EWBG in the $\lambda_{\Phi}$--$m_{\Phi}$ plane for $\lambda_{\Phi H} = 0.1$
(left) and $\lambda_{\Phi H} = 0.2$ (right).  The purple triangles mark points
where the first transition in the $\phi$ direction is strongly first-order and
the second is second- or weakly first-order; the red dots correspond to both transitions being 
strongly first-order, with the second transition completing efficiently. The
solid blue lines show the boundaries of the wedge region discussed in
the text, while the colour shading and gray dashed contours show the estimated strength of the first
transition based on the $w_c/T_c^w$ criterion. To the left of the diagonal black line, two minima exist at tree-level.
Above the horizontal black line, the standard Higgs vacuum is unstable at $T=0$. This
region is not visible in the parameter space shown for $\lambda_{\Phi H}=0.2$.
}
\end{figure}

  In Fig.~\ref{fig:sterile} we show the main results of our parameter scans 
in the $\lambda_{\Phi}$--$m_{\Phi}$ plane for 
$\lambda_{\Phi H}= 0.1$ (left) and $\lambda_{\Phi H}=0.2$ (right).
The solid red and purple dots correspond to specific parameter points that give
rise to a two-stage electroweak phase transition that is potentially
viable for EWBG.  For these points, the first transition occurs in
the $\phi$ direction and is strongly first-order with
\beq
w_n/T_n^w \geq 1 \ ,
\eeq
where $T_n^w$ is the nucleation temperature computed 
with \texttt{CosmoTransitions}, and $w_n$ is the VEV of $\phi$ 
in the symmetry-breaking vacuum when this occurs.  
Among the parameter points where the first transition is strongly
first-order, we also demand that the second transition completes
efficiently. As discussed above, first-order transitions complete when the bubble
nucleation rate overcomes the expansion rate of the Universe, which
for thermal tunneling processes roughly
takes place when $S_3/T_n<140$. The other features of Fig.~\ref{fig:sterile} 
will be explained below.

  The solid blue and dashed black lines in Fig.~\ref{fig:sterile}
correspond to the conditions of Eqs.~(\ref{eq:toy1}--\ref{eq:toy4})
derived for the toy model when applied to this theory.
To translate these conditions to the inert doublet model, we note that
the tree-level potential of Eq.~\eqref{eq:2HDMsterile}
projected onto the neutral real scalar components of $H$ and $\Phi$
coincides with the (tree-level) toy model presented in Section~\ref{sec:toy}.
The specific parameter identifications are
\beq
\begin{array}{lcrlcrlcr}
m_H^2 &\to&\mu_1^2,~~~&m_{\Phi}^2&\to&\mu_2^2,~~~\\
\lambda &\to& \lambda_1,~~~&\lambda_{\Phi} &\to& \lambda_2,~~~
&(\lambda_{\Phi H}+\tilde\lambda_{\Phi H}) ~\to~ \lambda_3,
\end{array}
\eeq
and the $a$ parameters of Eq.~\eqref{eq:a} now include gauge boson contributions. 
As in Fig.~\ref{fig:toy}, the top of the wedge region enclosed by
the thick solid blue lines bounds the region where the standard Higgs
minimum is deeper at $T=0$. Above the lower boundary of the wedge
the $\phi$ direction is destabilized before the $h$ direction
to leading order in the thermal expansion of the potential.
The upper horizontal black line in the left panel ($\lambda_{\Phi H} = 0.1$)
shows where the standard vacuum disappears entirely, 
while to the right of the black diagonal line the standard
Higgs vacuum is the only one present at $T=0$.  The horizontal line
is also present for $\lambda_{\Phi H} = 0.2$, but it is not visible
within the parameter ranges shown.  

  The shading and gray dashed contours in Fig.~\ref{fig:sterile} 
correspond to $w_c/T_c^w$, defined to be the ratio of the $\phi$ VEV 
at the local extremum in the $(0,\phi)$ direction to the critical temperature $T_c^w$ 
(at which $(0,\phi)$ becomes degenerate with the origin). 
Note that this definition does not require that the extrema at
$(0,w_c)$ or the origin be local minima, and thus the ratio $w_c/T_c^w$
does not necessarily correspond to a physical phase transition
as it is usually applied.  However, such an interpretation can be
made within the wedge region bounded by the solid blue lines discussed above.
Within the wedge, we also show contours of $v_c/T_c^v$ as solid green lines, 
defined similarly to be the ratio of the $h$ VEV at the local extremum 
$(v,w^\prime)$ at the critical temperature $T_c^v$ (when it becomes degenerate with
the local minimum at $(0,w)$). 

\begin{figure}[ttt]
\centering
 \includegraphics[scale=.5]{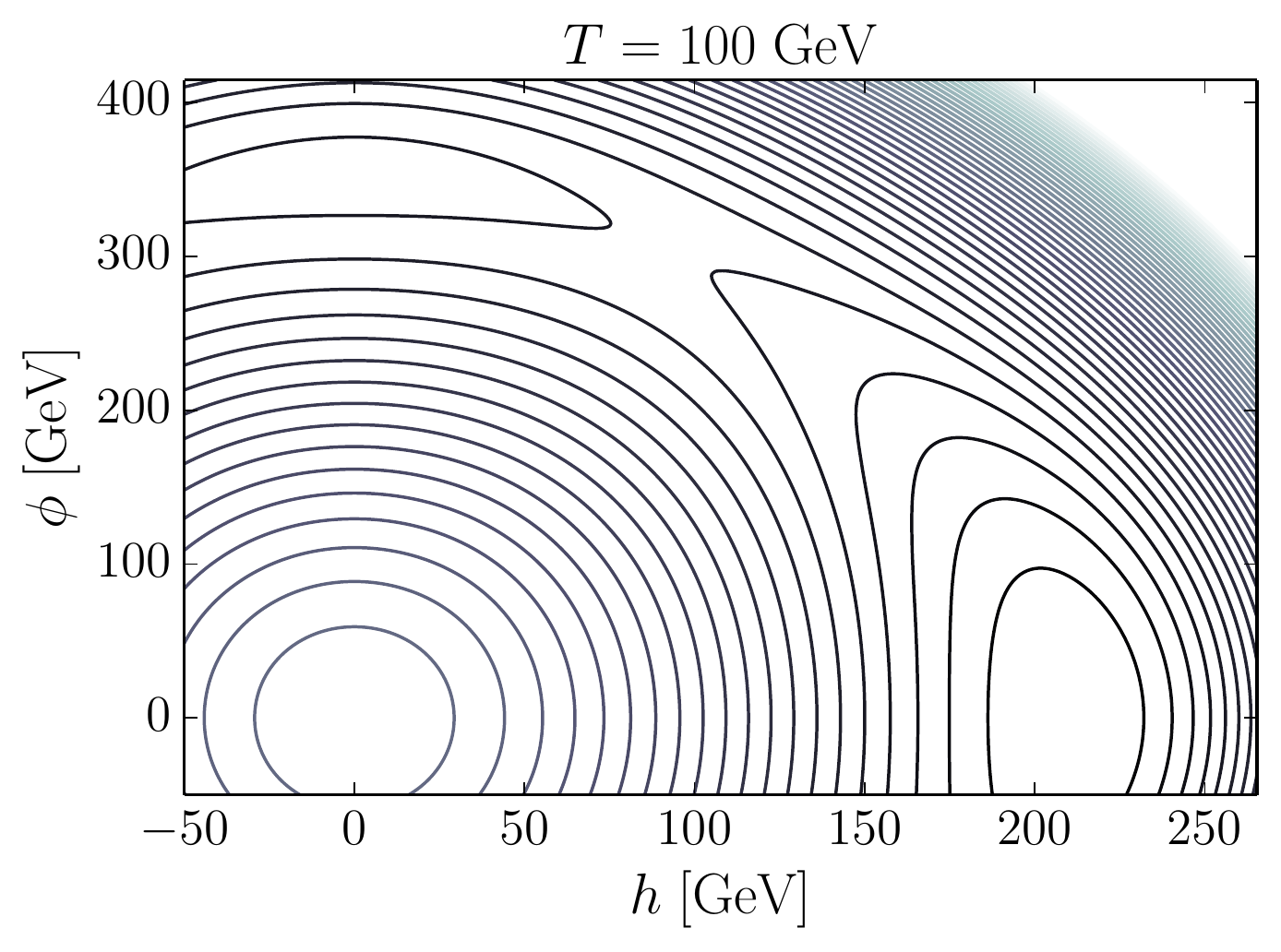}
  \includegraphics[scale=.5]{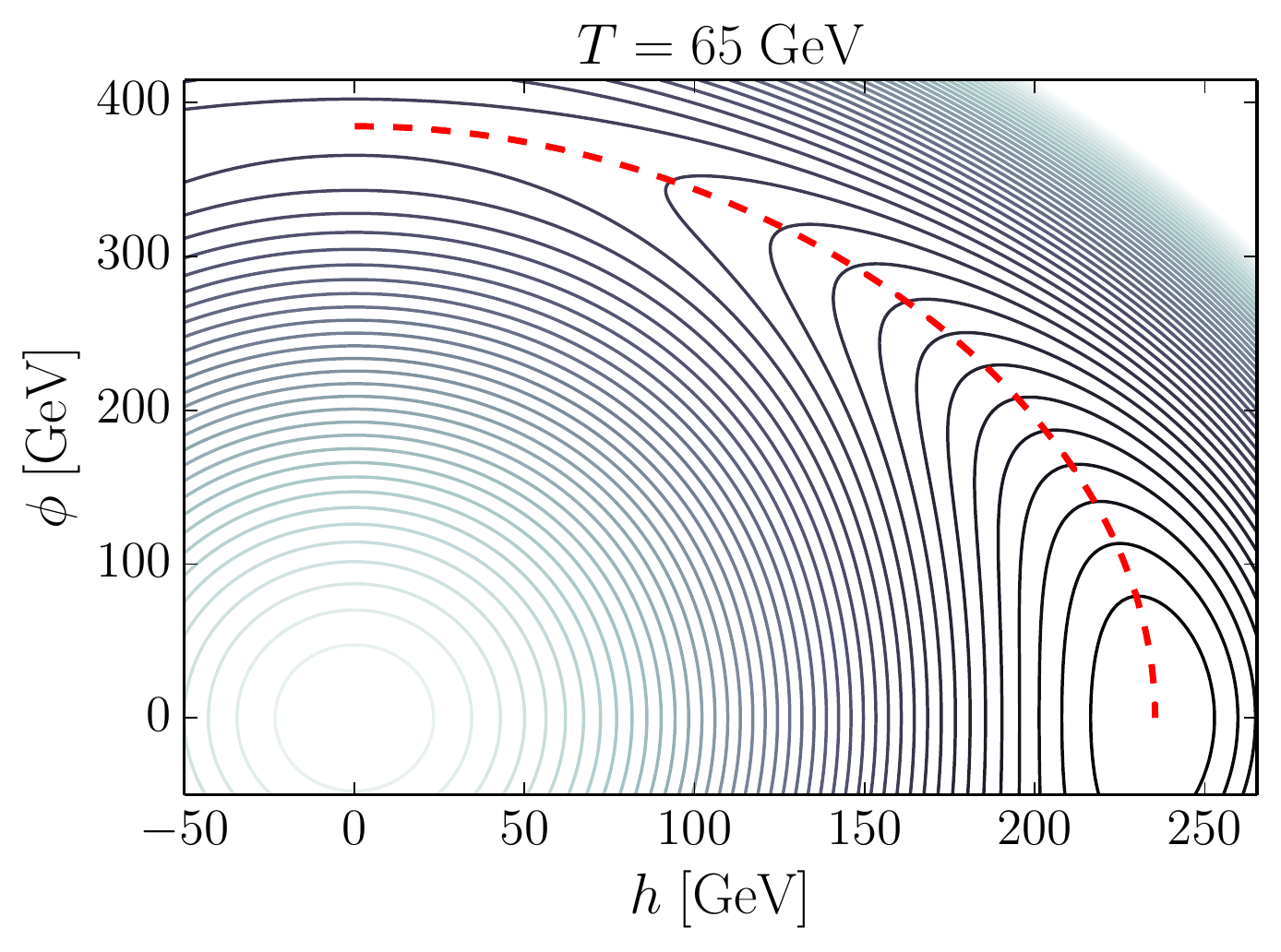}
\caption{
Contour plot of the one-loop effective potential in the $\phi$--$h$
plane of the inert doublet model for $\lambda_{\Phi H}=0.1$,
$\tilde\lambda_{\Phi H}=-0.001$, $\lambda_{\Phi}=0.0285$,
$m^2_{\Phi}=(45.3\,\gev)^2$, at $T=100$ GeV (left) and $T= 65$ GeV (right).
Two local minima are present at $T=100\,\gev$ while only the standard
electroweak minimum is stable at $T=65\,\gev$.  The dashed red line shows the
trajectory of the second order transition from the exotic to the SM Higgs vacuum.
\label{fig:valley} }
\end{figure}

  The EWBG-viable points in Fig.~\ref{fig:sterile} are also colour coded
according to whether the second phase transition is strongly first-order (red), 
or second-order or weakly first-order (purple).  For the red 
points, we have checked that the second transition completes efficiently
using \texttt{CosmoTransitions}.  The second transition in this case 
will inject some entropy into the cosmological plasma and dilute the
the baryon asymmetry produced at the first transition.  We have verified that 
this entropy injection is small for all of the points considered, typically
resulting in a percent-level dilution of the asymmetry. 
For the purple points, we have checked that the 
combination $\sqrt{h^2+\phi^2}/T$ remains greater than unity
over the course of the transition to ensure a sufficient suppression of
sphaleron washout during the second transition~~\cite{Ahriche:2014jna}. 
An example of this situation is shown in Fig.~\ref{fig:valley}. 
The left panel of the figure shows that two minima exist away from 
the origin at high temperature $T=100\,\gev$.
In the right panel, we see that at a lower temperature of $T = 65\,\gev$, 
the $\phi\neq0$ minimum becomes a saddle point and the field rolls down 
to the $h\neq0$ vacuum following the field trajectory shown 
by the dashed red line.

  Note that the solid red and purple dots in
Fig.~\ref{fig:sterile} featuring a two-stage electroweak phase transition
suitable for EWBG are found only in the left 
$\lambda_{\Phi H} = 0.1$ panel, and all of them lie within
the wedge region and to the right of the diagonal dashed boundary line
where $\Delta\lambda_2 > 0$.  These features exemplify the general
expectations derived from the toy model of Section~\ref{sec:toy}.
To the left of the diagonal boundary, a tree-level barrier between
the local minima along the $\phi$ and $h$ directions persists at $T=0$,
and the second transition from the exotic to the standard electroweak
vacuum never completes.  The viable points also lie in the leftmost
part of this subset of the wedge region where $w_c/T_c^w \gtrsim 1$
since this ratio is often a good estimate for $w_n/T_n^w$. 
These two considerations also explain why no points are found for 
$\lambda_{\Phi H} =0.2$ (right panel of Fig.~\ref{fig:sterile}); 
the diagonal boundary for $\lambda_2 >0$ now lies
further off to the right where $w_c/T_c^w$ is very small, and it is
not possible for the first transition to be strongly first-order,
while also having the second transition complete efficiently.

More generally, we only find two-stage electroweak phase transitions
suitable for EWBG in a limited range $\lambda_{\Phi H} \sim 0.1$.
Significantly larger values face the same challenge as $\lambda_{\Phi H} = 0.2$,
with the second transition not completing efficiently.
Smaller values of $\lambda_{\Phi H} \lesssim 0.07$ are also problematic
because they destabilize the standard Higgs minimum in the parameter region
where the exotic direction is destabilized first
(corresponding to lowering the horizontal black line in Fig.~\ref{fig:sterile}
related to the condition of Eq.~\eqref{eq:toy1} below the
lower boundary of the wedge).
We expect that a wider range of $\lambda_{\Phi H}$ is possible
with additional matter coupling to $H$ or $\Phi$.

\subsection{Comments on Gauge Dependence} \label{sec:gaugedep}

  The viable points in Fig.~\ref{fig:sterile} rely primarily
on the non-analytic gauge boson contributions to the thermal 
effective potential to drive the first transition in the $\phi$ 
direction to be first-order.
The challenge presented by these contributions is that they 
are gauge-dependent~\cite{Patel:2011th,Garny:2012cg}, and they 
lead to a gauge dependence in the ratio $w_n/T_n^w$ used to 
estimate the strength of the first phase transition.
For this reason, some care must be taken in interpreting the results 
of the calculation, which were obtained in the Landau gauge ($\xi=0$).
  
  To check the robustness of our findings, we have also analyzed
the transition strength of several scan points following the
gauge-invariant prescription of Ref.~\cite{Patel:2011th} to first-order
in the loop expansion.  While the critical temperatures change substantially
(reflecting the breakdown of the loop expansion near $T_c$), 
we find that viable two-step
transitions still typically occur for the points shown, although usually
with a reduced strength.  Given the significant uncertainty in 
determining the rate of non-perturbative baryon washout 
in the broken phase~\cite{Patel:2011th}, these points remain very plausible
candidates for two-stage EWBG.

  Let us also mention that the effective potential for $\phi$ along
the $h=0$ axis closely mirrors the potential for the SM Higgs 
in the limit $m_h \ll m_W$, which can be similarly driven strongly first-order
by gauge boson loop effects.  This scenario has been studied in 
great detail, with the literature including a number of non-perturbative 
studies~\cite{Kajantie:1995kf,Kajantie:1996mn} which do not suffer from gauge-dependence.  The strength of
the $\phi$ transition we find matches reasonably well with those reported in these previous investigations,
giving us further confidence in the reliability of our results.

\subsection{Comments on a Triplet Extension\label{subsec:triplet}}

To conclude this section, we review the results for the real triplet 
extension of the Standard Model studied in Ref.~\cite{Patel:2012pi}
and we compare them to the inert doublet model discussed above. 
The triplet does not couple to the SM fermions, thereby allowing 
the Higgs to receive stronger thermal corrections than the exotic states, 
in line with the general requirements for viable models with two-stage 
transitions delineated in Section~\ref{subsec:general2scalars}. However, unlike
the inert doublet case, the triplet also couples more strongly than 
the Higgs to the electroweak gauge bosons and thus has 
a relatively
enhanced contribution to its thermal mass from these states.
Hence, we expect a more limited parameter space for which two strongly 
first-order transitions are feasible relative to the models 
we have considered thus far.

The model considered in Ref.~\cite{Patel:2012pi} consists of the SM
plus a real $SU(2)_L$ triplet, $\Sigma=(\Sigma^1,\Sigma^2,\Sigma^3)$, with zero hypercharge $Y=0$ and the relevant interactions given by
\begin{align*}
V(H,\Sigma) =  -m^2_HH^\dagger H-\frac{1}{2}m^2_\Sigma\Sigma\ccdot\Sigma+\frac{\lambda}{2}(H^\dagger H)^2+\frac{\lambda_\Sigma}{4}(\Sigma\ccdot\Sigma)^2+\frac{\lambda_{\Sigma H}}{2}(H^\dagger H)(\Sigma\ccdot\Sigma).
\end{align*}
Projecting these fields onto their real neutral components, this potential
can be matched to the toy model of Section~\ref{sec:toy} with the identifications
\beq
\begin{array}{lcrlcrlcr}
m_H^2 &\to&\mu_1^2,~~~&m_{\Sigma}^2&\to&\mu_2^2,~~~\\
\lambda &\to& \lambda_1,~~~&\lambda_{\Sigma} &\to& \lambda_2/2,~~~
&(\lambda_{\Phi H}+\tilde\lambda_{\Phi H}) ~\to~ \lambda_3 \ .
\end{array}
\eeq
The thermal mass corrections must also be modified appropriately
and can be easily deduced from Ref.~\cite{Patel:2012pi}.

\begin{figure}[ttt]
\centering
\includegraphics[width=0.47\textwidth]{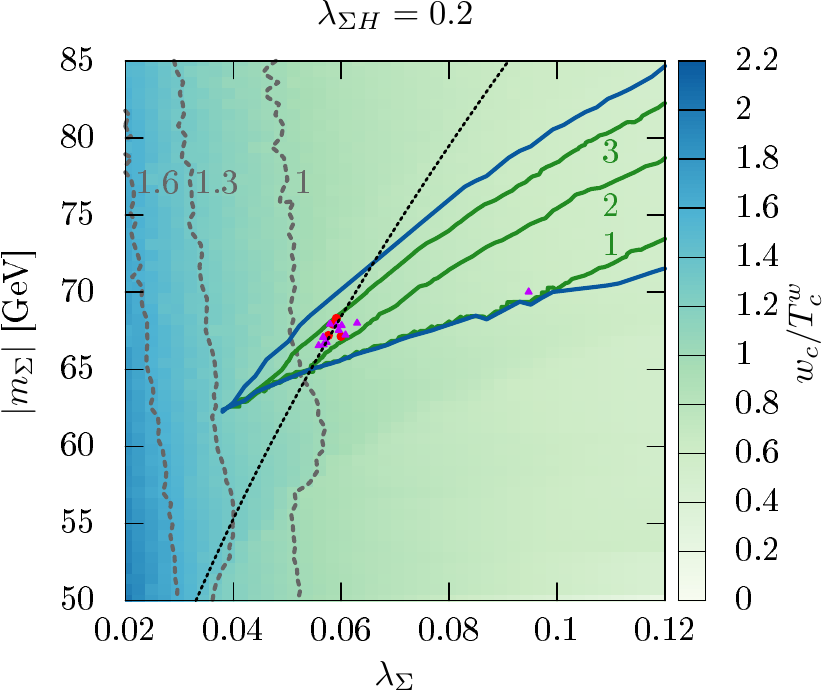}
\caption{\label{fig:triplet}
Parameter points leading to a two-stage electroweak phase transition amenable
to EWBG in the $\lambda_{\Sigma}$--$m_{\Sigma}$ plane for 
the real triplet extension of the SM with $\lambda_{\Sigma H} = 0.2$. 
The purple triangles mark points
where the first transition in the $\phi$ direction is strongly first-order and
the second is second- or weakly first-order; the red dots correspond to 
both transitions being strongly first-order, with the second transition 
completing efficiently.  The solid blue lines show the boundaries 
of the wedge region discussed in Section~\ref{subsec:sterile},
while the colour shading and gray dashed contours show the 
estimated strength of the first transition based on the 
$w_c/T_c^w$ criterion. Points to the left of the diagonal 
line exhibit two minima separated by a barrier at $T=0$.
}
\end{figure}

We have investigated this theory using the same techniques as for 
the inert doublet model by scanning over the parameters 
$\lambda_\Sigma$ and $m^2_\Sigma$ with fixed $\lambda_{\Sigma H}\in\{0.1,0.3\}$.
Figure~\ref{fig:triplet} shows the results of this scan 
for $\lambda_{\Sigma H} = 0.2$, with the plot format directly
analogous to Fig.~\ref{fig:sterile}.  As before, the purple triangles
and red dots correspond to parameter points that are suitable for EWBG
where the first phase transition in the exotic $\Sigma$ direction 
is strongly first-order, with the second transition being strongly 
first-order (purple triangles), or second-order or weakly first-order (red dots).

  These results show many similarities to the findings for the inert
doublet model and are consistent with the toy model expectations.
As before, the potentially viable points all lie within the
wedge region bounded by the solid blue lines.   They also lie
mostly to the right of the dotted black diagonal line where no
exotic minimum is present at zero temperature.
In general, we also find fewer points in the triplet case that are
suitable for two-stage EWBG, as well as larger allowed values for 
$\lambda_{\Sigma H}$ and $m_{\Sigma}^2$.  This can be understood
in terms of the stronger stabilization of the triplet direction
from thermal gauge boson loops.  To destabilize the triplet direction
before the Higgs, a larger $m_{\Sigma}^2$ is needed, 
which requires in turn a larger $\lambda_{\Sigma H}$ to drive 
the transition first-order.

  The results of our scan also appear to be consistent with the findings 
of Ref.~\cite{Patel:2012pi}, although we also focus on a slightly
different region of the parameter space.  Relative to that work,
we impose a stronger requirement on the strength of the first
phase transition in the $\Sigma$ direction, although the weaker
transitions considered in Ref.~\cite{Patel:2012pi} are potentially
also consistent with EWBG given the significant uncertainties
associated with determining the extent of baryon washout by non-perturbative
transitions~\cite{Patel:2011th}.  
After the first transition in this model, baryon washout
can occur both through sphalerons, and the scattering 
of baryons with the $SU(2)_L$ monopoles that can arise 
in the broken-triplet phase~\cite{Patel:2012pi}. 
The approximate condition for the suppression of sphalerons is
similar to the usual electroweak doublet, with $w_n/T_n \gtrsim 0.86$
at leading order~\cite{Ahriche:2014jna}.  The corresponding requirement 
for monopole suppression is more uncertain, but it is argued in
Ref.~\cite{Patel:2012pi} to be similar to that for sphaleron suppression.
With our stronger condition of $w_n/T_n^w \geq 1$, 
the coupling value $\lambda_{\Sigma H} = 0.2$
is about as large as possible.

\section{Phenomenological Constraints and the Need for More}
\label{sec:constraints}

  In the previous section, we showed that two-stage phase transitions 
suitable for EWBG can occur in models with an additional inert
Higgs doublet.  We turn next to investigate whether
the model parameter space of these scenarios can be consistent
with existing experimental constraints.  In general, we find
that all the parameter points that exhibit two-stage phase transitions are 
in conflict with observation due to the existence of light inert states 
in the spectrum.  To correct this, we construct a mild singlet extension of the
inert doublet model that drives up the inert doublet mass without significantly
modifying the two-stage electroweak phase transition.

\subsection{Mass Spectrum of the Inert Model}

  The inert doublet model presented above has four new physical degrees 
of freedom, in addition to the usual SM states. 
They consist of neutral and charged complex scalars, $\phi^0$ and $\phi^{\pm}$. 
At tree-level, their masses are
\begin{align}
m_{\phi^0}^2&=\frac{1}{2}\left(\lambda_{\Phi H}+\widetilde{\lambda}_{\Phi H}\right)v^2-m_{\Phi}^2\\
m_{\phi^{\pm}}^2&=\frac{1}{2}\lambda_{\Phi H}v^2 - m_{\Phi}^2
\end{align}
where $v=246\,\gev$.  The small and slightly negative 
$\tilde{\lambda}_{\Phi H}$ used in the analysis above implies that the
charged states are slightly heavier than the neutral ones.
Note as well that a non-zero $\lambda_5$ coupling in 
Eq.~\eqref{eq:2HDMsterile} would split the complex $\phi^0$ state
into a scalar $S$ and a pseudoscalar $A$ separated in mass 
by $\Delta m^2 = \lambda_5v^2$.  We will focus on the $\lambda_5\to 0$ limit, 
but we will comment on those cases where a small non-zero value 
can significantly modify the phenomenology.

  Comparing these expressions to the parameter values found to yield
two-stage transitions, all the points that are consistent with 
EWBG lead to new scalars that are very light,
$m_{\phi^{0,\pm}} \lesssim 35\,\gev$.  We expect this to hold beyond
the specific parameter ranges studied above.  Larger scalar masses
would require larger values of $\lambda_{\Phi H}$ and smaller values
of $m_{\Phi}^2$.  However, increasing $\lambda_{\Phi H}$ makes it more
difficult for the second transition to complete (corresponding
to Eq.~\eqref{eq:toy4} with $\lambda_3\to\lambda_{\Phi H}$),
while decreasing $m_{\Phi}^2$ tends to destabilize the standard Higgs $h$
direction before the exotic $\phi$ direction
(corresponding to Eq.~\eqref{eq:toy3} with $\mu_2^2 \to m_{\Phi}^2$).  

  In the analysis to follow, we will see that inert scalar masses
below about $63\,\gev$ are inconsistent with current experimental bounds.
With this in mind, we will study the limits on the theory while 
allowing for an additional contribution to the scalar masses, 
which we will parametrize according to
\beq
m^2_{\phi^{0,\pm}} \to m^2_{\phi^{0,\pm}} + \Delta^2 \ .
\eeq
We will return later to exhibit a singlet extension of the
inert doublet model that provides such an enhancement without significantly
changing the phase transition structure or the other direct constraints on
the theory.

\subsection{Precision Electroweak Tests}

  The states from the new electroweak doublet can modify precision 
electroweak observables through their \emph{oblique} loop corrections 
to the weak vector bosons~~\cite{Peskin:1990zt,Peskin:1991sw}.
Explicit expressions for these effects in the inert doublet model can be 
found in Ref.~\cite{Barbieri:2006dq}, and the most important constraint 
usually comes from $\Delta T$.  Evaluating them, we find only 
very mild shifts in the oblique parameters that are consistent
with current limits even for $\Delta^2 = 0$.  This can be 
understood from the custodial $SU(2)$ of the theory
in the limit of $\tilde\lambda_{\Phi H} \to 0$~\cite{Hambye:2007vf}, 
which is realized to a good approximation for the small value 
$\tilde\lambda_{\Phi H} = -0.001$ chosen in our analysis.

  The new scalar states can alter the behaviour of the $W$ and $Z$
vector bosons in a more direct way by opening up new decay channels.
To be consistent with data, new vector boson decay channels must be
nearly or completely kinematically forbidden, implying 
\beq
2m_{\phi^0},\;2m_{\phi^\pm} \gtrsim m_Z \ ,~~~~~~~m_{\phi^0}+m_{\phi^{\pm}} \gtrsim m_W \ .
\eeq
The first limit comes from the invisible decay width 
of the $Z$, or from searches for 
$Z\to f\bar{f}+\nu\bar{\nu}$.~\cite{Cao:2007rm,Dolle:2009fn}.
The second limit above follows from the precisely measured decay 
width of the $W$~\cite{Agashe:2014kda}.

\subsection{Direct Collider Searches}

   Searches for supersymmetry at LEP can be applied to derive limits
on the new scalars.  A reinterpretation of the OPAL search 
for charginos~\cite{Abbiendi:2003ji,Abbiendi:2003sc} 
suggests $m_{H^{\pm}}\gtrsim 70 \hspace{0.1cm} {\rm GeV}$~\cite{Pierce:2007ut},
although this limit disappears for mass splittings between the
charged and neutral states below about $5\,\gev$.
Similarly, the LEP search for neutralinos ($\chi_1^0\chi_2^0$)
can also be recast into a limit on the production cross-sections for the 
neutral inert scalar $\phi^0$.  If the mass splitting between 
the scalar and pseudoscalar components of $\phi^0$ 
is at least $\Delta m > 8\,\gev$, the bound is
$m_{\phi^0} \gtrsim 100 \hspace{0.1cm} {\rm GeV}$~\cite{Lundstrom:2008ai}.
However, for the small mass splittings $\Delta m < 8\,\gev$
that arise in the $\lambda_5\to 0$ limit considered here, this
bound disappears and the only relevant LEP limit on the neutral state 
is $m_{\phi^0} \gtrsim m_Z/2$.

  Data from the LHC has also been applied to constrain new inert doublets.
These analyses typically use many of the same methods 
as for chargino and neutralino 
searches~\cite{Cao:2007rm,Dolle:2009fn,Dolle:2009ft,Miao:2010rg,Gustafsson:2012aj,Belanger:2015kga}.
New bounds have been obtained beyond LEP, but they do not provide a useful
constraint on the model parameter space in the nearly-degenerate limit 
considered here~\cite{Dolle:2009ft,Miao:2010rg,Gustafsson:2012aj,Belanger:2015kga,Arhrib:2013ela}.

\subsection{Higgs Boson Decays}

  The presence of new electroweakly-charged scalars can have a significant
effect on how the SM-like Higgs boson decays.  These effects can be
direct if the Higgs decays into pairs of the new states, or they can
be indirect from the new scalars running in loops.  
We examine both possibilities here.

  Direct decays of the Higgs into the scalars can occur
if they are light enough.  The decay width to $\phi^0$ is~\cite{Pierce:2007ut}
\beq
\Gamma(h\to \phi^0\phi^0) = 
\frac{(\lambda_{\Phi H}+\tilde{\lambda}_{\Phi H})^2}{16\pi}\frac{v^2}{m_h}
\sqrt{1-\lrf{2m_{\phi}}{m_h}^2} \ .
\eeq
Away from the kinematic threshold $m_h = 2m_{\phi^0}$, 
this mode (and possibly its charged counterpart) completely  
dominates the Higgs branching for 
$(\lambda_{\Phi H}+\tilde{\lambda}_{\Phi H}) \gtrsim 0.05$~\cite{Arhrib:2013ela},
and is constrained by current Higgs data to lie below
$(\lambda_{\Phi H}+\tilde{\lambda}_{\Phi H}) \lesssim 7\times 10^{-3}$~\cite{
Belanger:2013xza}.  Larger values of this coupling appear to be needed 
for a viable two-stage electroweak phase transition, 
so we will focus on masses above $m_{\phi^0} \gtrsim 63\,\gev$
where such decays are kinematically forbidden (on-shell).

\begin{figure}[ttt]
\centering
\includegraphics[width=0.47\textwidth]{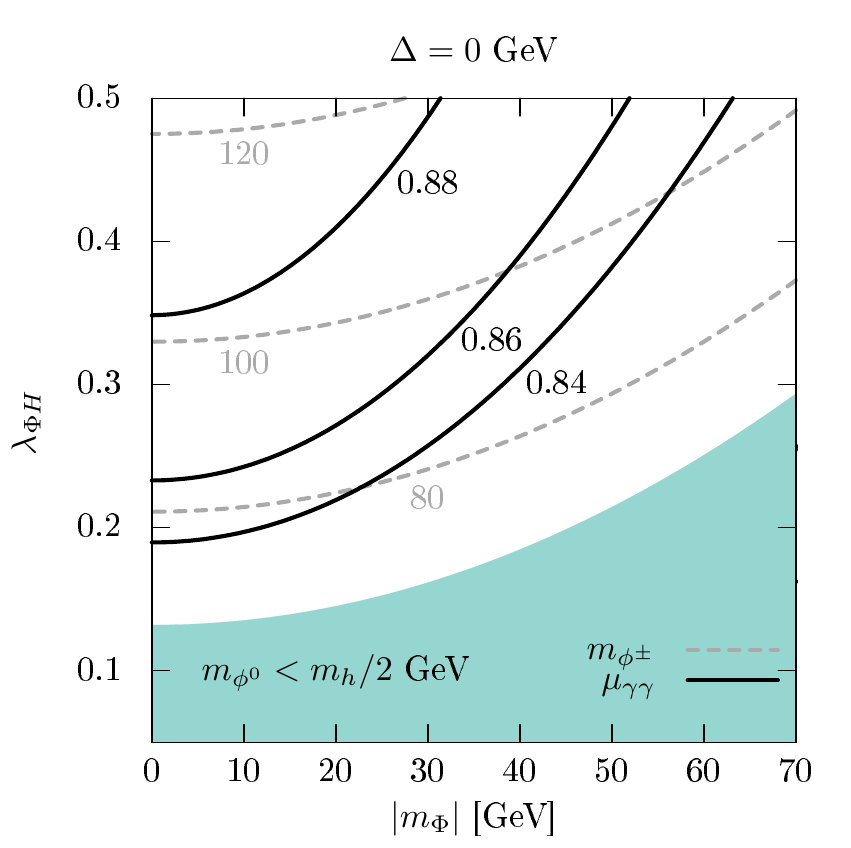}
\includegraphics[width=0.47\textwidth]{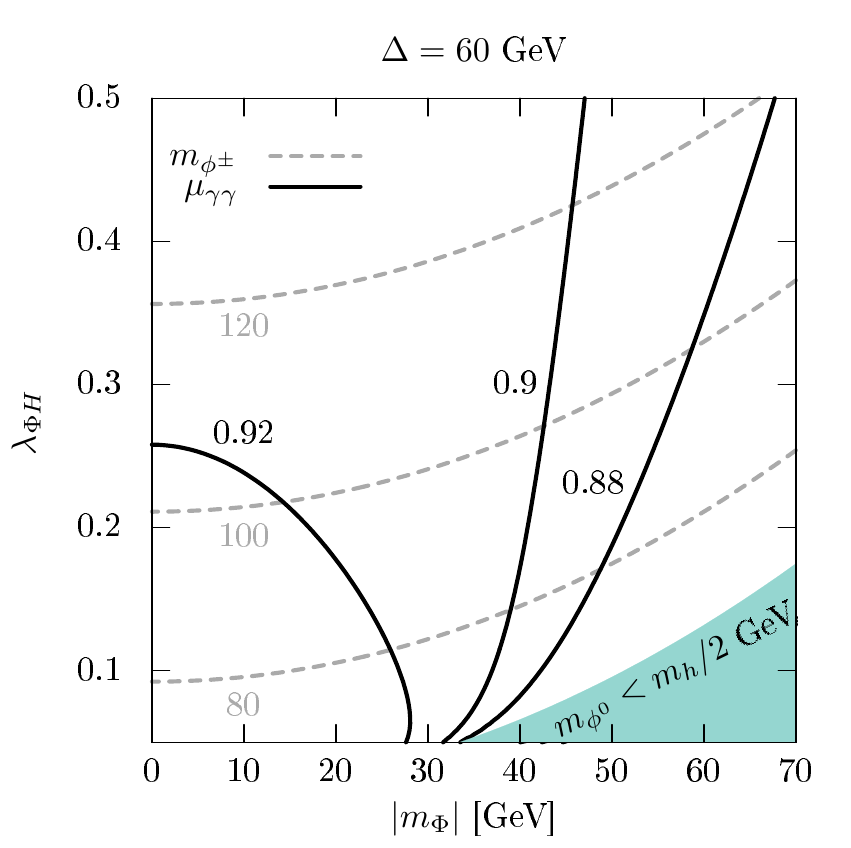}
\caption{\label{fig:inert2hdm_mugamgam} Contours showing deviations in the 
Higgs to diphoton signal rate $\mu_{\gamma\gamma}$ (solid black lines)
and the charged scalar mass $m_{\phi^{\pm}}$ (dashed gray lines) 
in the $m_{\Phi}$--$\lambda_{\Phi H}$ plane of the inert doublet model
with the mass correction parameter equal to 
$\Delta=0$ (left) and $\Delta=60\,\gev$ (right).
}
\end{figure}

  If the direct decays are closed, the most important effect of the new
doublet on the SM-like Higgs is to modify its decay width to
$\gamma \gamma$ and $\gamma Z$~\cite{Swiezewska:2012eh,Arhrib:2012ia,Celis:2013rcs}.  
In particular, the $\phi^{\pm}$ states
will contribute a new loop to the amplitudes for these processes
at a level that is similar to the dominant $W$ loop in the SM.
The other Higgs decay modes relevant at the LHC will not be changed
at leading order.  Modifications to the branching ratios of the Higgs 
were computed using the formulae of Refs.~\cite{hhg,Branco:2011iw,Djouadi:2005gj,Spira:1995rr}, 
while the SM width was evaluated using \texttt{HDECAY}~\cite{Djouadi:1997yw}. 
The results for the diphoton channel are shown 
in Fig.~\ref{fig:inert2hdm_mugamgam}, where we show the deviation 
in the signal rate 
$\mu_{\gamma\gamma} = (\sigma\; \mathrm{BR})/(\sigma\; \mathrm{BR})_\mathrm{SM}$
in the $\lambda_{\Phi}$--$m_{\Phi}$ plane for $\Delta = 0$ (left) and 
$\Delta = 60\,\gev$ (right).  In both panels the deviations are
generally consistent with the signal strengths observed by 
at ATLAS~\cite{Aad:2014eha} and CMS~\cite{Khachatryan:2014ira},
\beq
\mu_{\gamma\gamma} = 1.17 \pm 0.27~~(\mathrm{ATLAS}) \ ,~~~~~
\mu_{\gamma\gamma} = 1.14^{+0.26}_{-0.23}~~(\mathrm{CMS}) \ .
\eeq  
Similar conclusions hold for the triplet case \cite{Patel:2012pi}.

\subsection{Dark Matter}

  The inert doublet model we have considered has an unbroken global 
$U(1)$ in the standard electroweak vacuum (in the $\lambda_5\to 0$ limit), 
and the lightest exotic state will therefore be stable and contribute 
to the density of dark matter.  
For $\tilde\lambda_{\Phi H} < 0$, this state will be the neutral scalar $\phi^0$.  
The relic abundance of the lightest scalar in inert doublet models
has been studied extensively in, \eg Refs.~\cite{LopezHonorez:2006gr,Dolle:2009fn,Gil:2012ya,Goudelis:2013uca,Arhrib:2013ela,Blinov:2015xx}, 
and we restrict ourselves to a few brief comments.

  For $|\tilde\lambda_{\Phi H}|\lesssim m_{\phi^0}^2/v^2$, 
the charged and neutral states are close in mass and produce 
a very small thermal relic density through 
coannihilation~\cite{Pierce:2007ut}.  As the neutral-charged mass splitting
increases (with a heavier charged state), the relic density increases as well 
but still remains well below the full dark matter density thanks to
efficient annihilation through the $Z^0$ to fermions~\cite{Barbieri:2006dq}.  
The relic density of the lightest state will increase further 
if the global $U(1)$ symmetry is broken to $\mathbb{Z}_2$ by introducing 
a small $\lambda_5$ coupling;
this separates the scalar and pseudoscalar components of $\phi^0$ and
moderates the annihilation via the $Z^0$~\cite{Barbieri:2006dq}.
Since the required splittings are small, we expect that all these possibilities
can be realized while maintaining a two-step electroweak phase transition
consistent with EWBG.

  The lightest inert scalar is also strongly constrained by direct searches
for dark matter, even if the scalar relic density is well below the full
dark matter density.  This is especially true when the scalar and pseudoscalar
components of $\phi^0$ are degenerate ($\lambda_5\to 0$), since there is an
extremely strong $\phi^0$-nucleon interaction mediated by $Z^0$ exchange.
When $\lambda_5$ is turned on, the $Z$ couples exclusively to 
the scalar--pseudoscalar combination and the scalar-nucleon scattering
is inelastic~\cite{Arina:2009um}.  Thus, it becomes irrelevant 
to direct detection searches for mass splittings above a few hundred keV.  
The next most important contribution to nucleon scattering comes from Higgs 
exchange~\cite{Barbieri:2006dq}, and even this is too large compared
to current limits from XENON100~\cite{Aprile:2012nq} 
and LUX~\cite{Akerib:2013tjd} for 
$(\lambda_{\Phi H}+\lambda_{\Phi H}+\lambda_5)/2 \gtrsim 0.02$ and
$m\in [10,100]\,\gev$ assuming a local scalar relic density 
of $\rho_{\phi} = 0.3\,\gev/\text{cm}^3$~\cite{Arhrib:2013ela}.  

  Taken together, these results suggest that it is unlikely for the 
lightest scalar to make up all of the dark matter while satisfying the 
limits from direct detection and producing a viable two-stage
electroweak transition.  On the other hand, the lightest scalar could make
up a small fraction of the dark matter density provided it is neutral with
at least a small mass splitting ($\gtrsim 100$\;keV) between its scalar
and pseudoscalar components.  Alternatively, the lightest scalar could
be destabilized by introducing an  explicit soft breaking of 
the $\mathbb{Z}_2$ symmetry~\cite{Enberg:2013jba,Enberg:2013ara}, 
allowing the would-be dark matter to decay.  If this breaking is small enough, 
the phase transition history is expected to be mostly unchanged.

\subsection{A Singlet Extension~\label{sec:singlet}}

  As discussed above, the exotic scalars $\phi^{0,\pm}$ are very light  
in the parameter regions found to yield a two-stage electroweak phase 
transition suitable for EWBG.  Such scalars are excluded by experimental
measurements of weak vector boson and Higgs decays.  We parametrized an
additional contribution to the scalar masses by a factor $\Delta$
such that $m_{\phi^{0,\pm}}^2 \to m_{\phi^{0,\pm}}^2 + \Delta^2$.  In this
section we exhibit a simple singlet extension of the theory to generate
such a $\Delta$ contribution that does not significantly disturb the
electroweak phase transitions.

  The potential in the extended theory for the scalar electroweak doublets 
and the new real singlet $S$ is taken to be
\begin{align}
\nonumber V(H,\Phi,S)~=~&-m^2_H H^\dagger H-m^2_\Phi \Phi^\dagger\Phi
+\frac{\lambda_H}{2}(H^\dagger H)^2+\frac{\lambda_\Phi}{2}(\Phi^\dagger \Phi)^2\\
&\label{eq:2HDMsinglet}+\lambda_{\Phi H}(H^\dagger H)(\Phi^\dagger \Phi)
+\tilde\lambda_{\Phi H}(\Phi^\dagger H)(H^\dagger \Phi) 
+ \left[\frac{1}{2}\lambda_5(H^{\dagger}\Phi)^2 + \hc\right]\\
&- \frac{1}{2}m^2_S S^2 +\frac{\lambda_S}{4!}S^4
+\frac{\lambda_{S H}}{2}H^\dagger H S^2
+\frac{\lambda_{S \Phi}}{2}\Phi^\dagger \Phi S^2
\nn
\ ,\nnmb
\end{align}
where possible additional terms can be forbidden with suitable symmetries. 
As before, we will invoke an approximate $U(1)$ symmetry to set 
$\lambda_5\rightarrow0$.
We will also set $\lambda_{SH} \to 0$ for now to avoid mixing with the SM Higgs,
but we will return to non-zero values below.
If the singlet develops a VEV, $S = v_S+\rho$, the masses 
of both the neutral and charged doublets $\phi^{0,\pm}$ will be
modified by $m_{\phi^{0,\pm}}^2 \to (m_{\phi^{0,\pm}}^2 + \Delta^2)$ with
\beq
\Delta^2 ~=~
\frac{\lambda_{S \Phi}}{2} v_{S}^2
~=~ \frac{3\lambda_{S \Phi}}{\lambda_{S}}
m_S^2
\ .
\eeq
The mass of the real singlet excitation in the present vacuum is
\beq
m_{\rho}^2 ~=~ \frac{\lambda_S}{3}v_S^2 
~=~ 2m_S^2 
\ .
\eeq

\begin{figure}[ttt]\centering
\includegraphics[width=8cm]{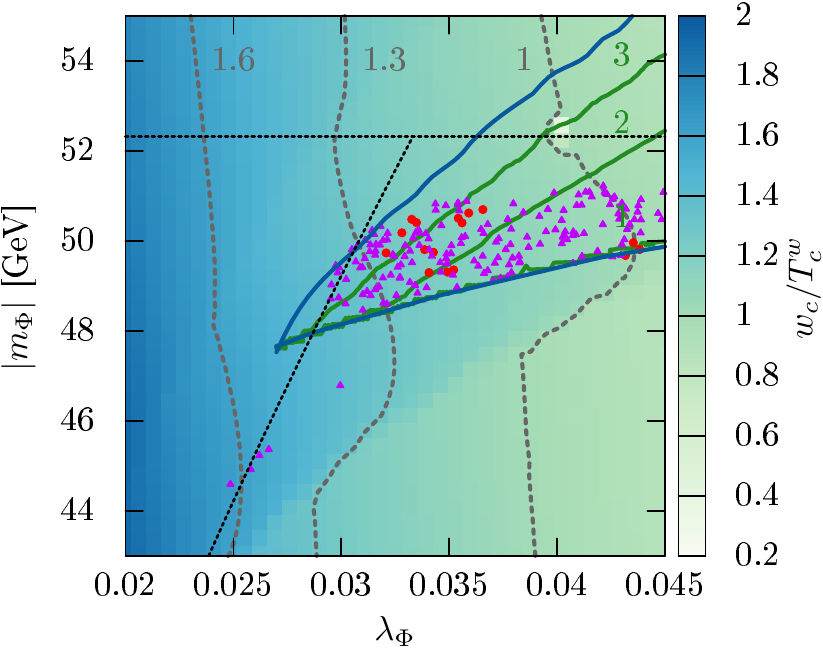}
\caption{\label{fig:scalar2}
Parameter points leading to a two-stage electroweak phase transition amenable
to EWBG in the $\lambda_{\Phi}$--$m_{\Phi}$ plane for the 
inert doublet model extended by a singlet $\lambda_{\Phi H} = 0.1$.  
The remaining parameters are chosen as described in the text. 
The singlet field $S$ develops a nonzero VEV at $T_c\sim20$
GeV, well below the temperature of the transitions in the figure. 
The various contours and solid points are the same as in Figs.~\ref{fig:sterile} and \ref{fig:triplet}.}
\end{figure}

  In Fig.~\ref{fig:scalar2} we show parameter points in this extended theory
that realize a two-step electroweak phase transition that is amenable to EWBG
for the model parameters $\lambda_{\Phi H} = 0.1$,
$\lambda_{S} = 0.001$, $m_S = 4.63\,\gev$, $\lambda_{S\Phi} = 0.72$,
and $\lambda_{SH} = 0$.  For these points, the $S$ field is found to
develop a VEV at temperatures near $T_c^S \sim 20\,\gev$,
well below the temperatures of the other transitions.  Thus,
the singlet mostly decouples from the dynamics of the electroweak
transitions for our choice of parameters, and its primary physical effect is to
generate the mass contribution $\Delta^2$ at late times.  

  The distribution of EWBG-viable points in Fig.~\ref{fig:scalar2}
is mostly similar to what was found for the pure inert doublet
(Fig.~\ref{fig:sterile}) and triplet (Fig.~\ref{fig:triplet}),
with most of the points in the wedge region to the right of the 
dashed black diagonal line.
However, there is also a small population of viable dots in the lower left 
of Fig.~\ref{fig:scalar2} outside the wedge region.
These points lie below the bottom of the wedge where the $h$ direction
is expected to be destabilized before the $\phi$ direction
based on the simple analysis of Section~\ref{sec:toy}.  The more complete 
analysis of the thermal evolution of the potential undertaken
here shows that some points can avoid this criterion to an extent.
Note as well that this second population, as well as a subset of points 
within the wedge, lie very close to, but to the left of the boundary 
where $\Delta\lambda_2 = 0$.  For these points, a local minimum
persists in the $\phi$ direction at zero temperature, but the barrier separating
it from the standard electroweak vacuum may be small enough for tunneling to
occur at an acceptable rate. The corresponding transitions would be weakly first-order.

  While we have set $\lambda_{SH}=0$ in the analysis above, this limit is
unnatural since loops will generate
\beq 
\Delta\lambda_{SH}(\mu) ~\sim~ 
\frac{\lambda_{S\Phi}\lambda_{\Phi H}}{(4\pi)^2}\ln\lrf{\mu}{\mu_0} \ ,
\eeq
where $\mu$ is the running scale and $\mu_0$ is a reference scale 
where $\lambda_{SH}$ vanishes.  
Similarly, the singlet self-coupling receives contributions 
from quantum corrections 
on the order of $\Delta\lambda_S \sim \lambda_{S\Phi}^2\ln(\mu/\mu_0)/(4\pi)^2$,
so naturalness sets an approximate lower bound on $\lambda_S$ as well. 
The phase transition picture described above will be preserved as long
as $\lambda_S$ and $\lambda_{SH}$ are small, 
near the lower range consistent with naturalness.
However, one could also imagine larger values where the singlet plays a key
role in the other phase transitions.  

  A non-zero value of $\lambda_{SH}$ also gives rise to mixing between
the singlet and the SM-like Higgs boson.  This \emph{Higgs portal} coupling 
allows the Higgs to decay to pairs of light singlets, and it opens a new decay
channel for the singlet states~\cite{Schabinger:2005ei,Patt:2006fw}.  
For $m_{\rho} < m_{\phi}$, the dominant decay of the mostly-singlet state
proceeds through the Higgs portal and will mirror the branching fraction
a SM-like Higgs would have if $m_h = m_{\rho}$.   Searches for
rare $B$ decays imply very strong limits on $\lambda_{SH}$ for 
$m_{\rho} < m_B-m_K$ that typically rule out natural values 
of this coupling~\cite{Batell:2009di,Clarke:2013aya,Falkowski:2015iwa}.  
For higher masses, the limits become much weaker, 
coming from a combination of LEP searches for a light Higgs boson 
(typically above the $b\bar{b}$ 
threshold)~\cite{Clarke:2013aya,Falkowski:2015iwa}, 
rare $\Upsilon$ decays~\cite{Clarke:2013aya}, 
and low-mass dimuon searches at the LHC~\cite{Clarke:2013aya}.  
These typically allow for small but natural values of $\lambda_{SH}$.

  Finally, let us note that this theory has a global $\mathbb{Z}_2$ symmetry
under which $S\to -S$ that is spontaneously broken when $S$ gets a VEV.
As it stands, this will lead to the formation of cosmologically problematic
domain walls~\cite{Kibble:1980mv}.  These can be eliminated without significantly
modifying the electroweak phase transition structure by adding a very 
small explicit soft breaking of the $\mathbb{Z}_2$, or by promoting $S$
to a complex scalar charged under a hidden $U(1)_x$ gauge group.

\section{Conclusions}\label{sec:conc}

Electroweak baryogenesis provides a well-motivated explanation of the observed
baryon asymmetry that is in principle testable at present day experiments. As
such, it is important to understand its possible realizations and the extent to
which they can be probed.  
Indeed, many existing proposals for EWBG are strongly constrained
by precision Higgs measurements and searches for new sources of $CP$ violation.
In this study, we have focused on a realization of EWBG
in which baryon creation occurs in an initial
phase transition to an exotic vacuum exhibiting broken 
electroweak symmetry~\cite{Patel:2012pi}. A later
transition to a Standard Model-like Higgs vacuum then occurs at some lower
temperature. Naively, this allows for the new physics involved in strengthening
the electroweak phase transition 
to be decoupled from the properties of the SM-like Higgs boson
observed at the LHC. The results of our study suggest that this 
is not entirely true.

While two-stage electroweak phase transitions have been studied previously 
and applied to EWBG, our investigation is novel in several respects. 
We have presented a simplified model that makes clear 
the conditions required for two-stage electroweak symmetry breaking 
in general extensions of the SM Higgs sector. We used the intuition 
gained from this toy model to study two-stage phase transitions 
in inert two-Higgs-doublet models, with and without additional matter content. 
These scenarios can exhibit two-step phase transitions suitable for EWBG.
Relative to the triplet model studied in Ref.~\cite{Patel:2012pi},
the inert doublet model appears to have a larger parameter region
where this can occur, and it avoids the possibility of
additional baryon washout by monopoles after the first transition.

  A seemingly general result of our study is that two-stage electroweak 
phase transitions predict new light electroweakly-charged particles.  
In the portion of the inert doublet model parameter space 
consistent with two-stage EWBG, 
these new states are so light that they are ruled by direct collider
searches.  We illustrated a generic solution to this problem in the form 
of a singlet-induced mass term for the problematic states that can still 
allow for successful electroweak baryogenesis from a two-stage transition.
Although the singlet contribution to the sterile particle masses allows for
two-stage scenarios consistent with current experimental data, the models we
have considered here still generically feature potentially 
observable deviations in the rate of $h\rightarrow \gamma \gamma$ 
relative to the Standard Model due to loop contributions from the new 
exotic states.  This prediction seems difficult to avoid since increasing 
the singlet-inert mixed quartic coupling to push up the electroweak-preserving
masses of these states would also lead to larger thermal mass corrections
to the exotic multiplet.  For large couplings, the phase transition 
to the SM-like Higgs vacuum will then occur first.   
We expect this feature to be generic, holding beyond just the inert doublet 
(and triplet) scenarios considered here. 
This exemplifies the virtue of electroweak baryogenesis as a
concretely testable scenario and one that deserves to be studied thoroughly,
both theoretically and in experiments exploring the electroweak scale.

\section*{Acknowledgements}

The authors would like to thank Wei Chao, Grigory Ovanesyan,
Michael Ramsey-Musolf, Brian Shuve, and Peter Winslow for useful
discussions. We also thank Michael Ramsey-Musolf for comments
on this manuscript.
This research is supported by the National Science 
and Engineering Research Council of Canada~(NSERC). 
Research  at  the  Perimeter  Institute  is  supported  in  part  by  the
Government of Canada through Industry Canada, and by the Province 
of Ontario through the Ministry of Research and Information (MRI).
NB thanks UC Santa Cruz and the Santa Cruz Institute for Particle Physics 
for hospitality where part of this work was completed.  
DM thanks the Perimeter Institute for their hospitality during part 
of the completion of this work. 
CT acknowledges support of the Spanish Government 
through grant FPA2011-24568 (MICINN), and thanks TRIUMF, 
where the ideas for this work were first discussed.

\appendix


\section{One-Loop Beta Functions and Anomalous Dimensions\label{app:RG}}

In this appendix we reproduce the beta functions and anomalous dimensions
relevant for computing the renormalization group-improved effective potential. 
We follow the standard convention of defining the beta function of
a coupling $\kappa$ to be
\beq
\beta_\kappa=\mu\frac{d\kappa}{d\mu} \ ,
\eeq
where $\mu$ is the renormalization scale.
For a field $\varphi$ with an associated (unmixed) wavefunction 
renormalization factor $Z_{\varphi}$ we define the anomalous 
dimension to be
\beq
\gamma_\varphi=\frac{1}{2 Z_\varphi}\mu\frac{d Z_\varphi}{d\mu} \ .
\eeq
We computed these at two loops in the $\overline{\rm MS}$ scheme 
in dimensional regularization following the general results of 
Refs.~\cite{Machacek:1983tz,Machacek:1983fi,Machacek:1984zw,Luo:2002ti}.
In the formulae below, we only show the one-loop result for compactness.
Note as well that we use the GUT normalization for the hypercharge 
gauge coupling, $g_1=\sqrt{5/3}g'$.

The beta functions for the inert doublet plus singlet model defined 
in Section~\ref{sec:singlet}, in the enhanced-symmetry limit $\lambda_5=0$, are
\beq
 \beta_{g_1}&=&\beta^{\mathrm{ SM}}_{g_1}+\frac{1}{(4\pi)^2}\frac{g_1^3}{10},\\
  \nn\beta_{g_2}&=&\beta^{\mathrm{SM}}_{g_2}+\frac{1}{(4\pi)^2}\frac{g_2^3}{6},\\
 \nn \beta_{g_3}&=&\beta^{\mathrm{SM}}_{g_3},\\
  \nn\beta_\lambda&=&\beta^{\mathrm{SM}}_{\lambda_1}+\frac{1}{(4\pi)^2}\left[4 \lambda _{\Phi H} \tilde{\lambda }_{\Phi H}+2 \tilde{\lambda }_{\Phi H}^2+\lambda _{{S H}}^2+4 \lambda _{\Phi H}^2\right],\\
 \nn\beta_{m^2_H}&=&\beta^{\mathrm{SM}}_{m^2_H}+\frac{1}{(4\pi)^2}\left[m^2_\Phi \left(2 \tilde{\lambda }_{\Phi H}+4 \lambda _{\Phi H}\right)+\lambda _{{S H}} m_{S }^2\right],\\
 \nn\beta_{\lambda_{\Phi}}&=&\frac{1}{(4\pi)^2}\bigg[2 \tilde{\lambda }_{\Phi H}^2+4 \lambda _{\Phi H} \tilde{\lambda }_{\Phi H}+\left(-\frac{9}{5}  g_1^2-9 g_2^2\right) 
 \lambda _{\Phi }+\frac{27 g_1^4}{100}+\frac{9}{10} g_2^2 g_1^2+\frac{9 g_2^4}{4}+12 \lambda _{\Phi }^2\\
\nn &&+4 \lambda _{\Phi H}^2+\lambda _{S\Phi}^2\bigg],\\
\nn\beta_{m^2_\Phi}&=&\frac{1}{(4\pi)^2}\left[m_H^2 \left(2 \tilde{\lambda }_{\Phi H}+4 \lambda _{\Phi H}\right)+m_{\Phi }^2 \left(-\frac{9}{10}  g_1^2-\frac{9 g_2^2}{2}+6 \lambda _{\Phi }\right)+m_S^2 \lambda _{S\Phi }\right],\\
 \nn\beta_{\lambda_{\Phi H}}&=&\frac{1}{(4\pi)^2}\bigg[\lambda _\Phi \left(2 \tilde{\lambda }_{\Phi H}+6 \lambda _{\Phi H}\right)+2 \tilde{\lambda }_{\Phi H}^2+\lambda  \left(2 \tilde{\lambda }_{\Phi H}+6 \lambda _{\Phi H}\right)+\lambda _{\Phi H} \bigg(6 y_b^2-\frac{9}{5}  g_1^2-9 g_2^2\\
\nn &&+\!6 y_t^2\!+2 y_{\tau }^2\bigg)+\frac{27 g_1^4}{100}-\frac{9}{10} g_2^2 g_1^2+\frac{9 g_2^4}{4}+4 \lambda _{\Phi H}^2+\lambda _{SH} \lambda _{S\Phi}\bigg],\\
 \nn\beta_{\tilde\lambda_{\Phi H}}&=&\frac{1}{(4\pi)^2}\bigg[\tilde{\lambda }_{\Phi H} \left(6 y_b^2-\frac{9}{5}  g_1^2-9 g_2^2+8 \lambda _{\Phi H}+6 y_t^2+2 y_{\tau }^2\right)+2 \lambda _\Phi \tilde{\lambda }_{\Phi H}+4 \tilde{\lambda }_{\Phi H}^2+2 \lambda  \tilde{\lambda }_{\Phi H}\\
\nn& &+\frac{9}{5} g_1^2 g_2^2\bigg],\\
\nn\beta_{\lambda_{S}}&=&\frac{1}{(4\pi)^2}\left[3 \lambda _{S }^2+12 \lambda _{{S H}}^2+12 \lambda _{S\Phi }^2\right],\\
\nn\beta_{m^2_S}&=&\frac{1}{(4\pi)^2}\left[4 m_H^2 \lambda _{{S H}}\!+\lambda _{S } m_{S }^2+4 m_{\Phi }^2 \lambda _{S\Phi }\right],\\
 \nn\beta_{\lambda_{S H}}&=&\frac{1}{(4\pi)^2}\bigg[\lambda _{{S H}} \left(6 y_b^2-\frac{9}{10}  g_1^2-\frac{9 g_2^2}{2}+6 y_t^2+2 y_{\tau }^2\right)+\lambda _{S } \lambda _{{S H}}+4 \lambda _{{S H}}^2+6 \lambda  \lambda _{{S H}}\\
 \nn& &+2 \lambda _{S\Phi } \tilde{\lambda }_{\Phi H}+4 \lambda _{\Phi H} \lambda _{S\Phi }\bigg],\\
\nn\beta_{\lambda_{S\Phi}} &=&\frac{1}{(4\pi)^2}\left[2 \lambda _{SH} \tilde{\lambda }_{\Phi H}+\lambda _{S\Phi } \left(-\frac{9}{10}  g_1^2-\frac{9 g_2^2}{2}+6 \lambda _{\Phi }+\lambda _S\right)+4 \lambda _{\Phi H} \lambda _{SH}+4 \lambda _{S\Phi }^2 \right] \ .
\eeq
In the equations above, the $\beta_\delta^{\rm SM}$ denotes the beta function
of the corresponding coupling in the Standard Model. The above beta functions
may also be applied to the pure inert doublet model by setting $\lambda_S =
\lambda_{SH} = \lambda_{S\Phi} = 0 = m_S^2$.
The anomalous dimensions are
\beq
\nn{\gamma_H}&=&\beta^{\mathrm{SM}}_{\gamma_H},\\
{\gamma_\Phi}&=&\frac{1}{(4\pi)^2}\left[\!-\frac{9}{20}  g_1^2\!-\frac{9 g_2^2}{4}\right],\\
\nn\gamma_{S} &=&0.
\eeq

\section{One-Loop Thermal Masses\label{app:T}}

  The leading high-temperature contributions to the bosonic self-energies
are needed in order to perform a daisy resummation of the corrections due to
the bosonic $n=0$ Matsubara modes.  These have been derived in the SM
in Ref.~\cite{Carrington:1991hz}, while partial results for inert doublet
models are given in Refs.~\cite{Cline:1995dg,Gil:2012ya}.  We extend
these results to include an additional singlet as in Section~\ref{sec:singlet}.

  For the gauge fields, only the self-energies corresponding to longitudinal 
polarizations receive thermal corrections at leading order. 
Denoting these with $\Pi_L^W$ and $\Pi_L^B$ for the $SU(2)_L$
and $U(1)_Y$ gauge bosons, respectively, one has 
\beq
 \Pi_L^W =  2g_2^2 T^2 \ ,~~~~~~~
 \Pi_L^B = \frac{6}{5}g_1^2 T^2 \ .
\eeq
Taking these together with the zero-temperature results, 
the temperature-corrected mass eigenvalues for the longitudinal components 
of the gauge bosons in a general $(h,\phi)\neq (0,0)$ background are
\begin{align}
 m_L^W=&\frac{1}{4}g_2^2(h^2+\phi^2)+2g_2^2T^2,\\
 m_L^Z=&\frac{1}{8}\Big(g_2^2+\frac{3}{5}g_1^2\Big)(h^2+\phi^2)+g_2^2 T^2+\frac{3}{5}g_1^2T^2+\Delta_1,\\
 m_L^\gamma=&\frac{1}{8}\Big(g_2^2+\frac{3}{5}g_1^2\Big)(h^2+\phi^2)+g_2^2 T^2+\frac{3}{5}g_1^2T^2-\Delta_1,\\
 \Delta_1=&\frac{1}{40} \sqrt{\left(5 g_2^2+3 g_1^2\right)^2 \left(\phi^2+8 T^2+h^2\right)^2-960 g_2^2 g_1^2 T^2 \left(\phi^2+4 T^2+h^2\right)} \ .
\end{align}

The leading thermal self-energies for the scalar doublets and the singlet are
\begin{align}
 \Pi_{H^\dagger H} = & \left(\frac{1}{4}\lambda 
+ \frac{1}{6}\lambda_{\Phi H}  + \frac{1}{12}\tilde\lambda_{\Phi H}
+\frac{1}{24}\lambda_{S H}
+ \frac{1}{4}y_t^2  
+\frac{3}{16}g_2^2+\frac{3}{80}g_1^2\right)T^2,\\
 \Pi_{\Phi^\dagger \Phi}  = & \left(\frac{1}{4}\lambda_\Phi 
+ \frac{1}{6}\lambda_{\Phi H}+ \frac{1}{12}\tilde\lambda_{\Phi H}
+\frac{1}{24}\lambda_{S \Phi}
+\frac{3}{16}g_2^2+\frac{3}{80}g_1^2\right)T^2,\\
\Pi_{SS} = &\left(\frac{1}{24}\lambda_S +\frac{1}{6}\lambda_{S H}+\frac{1}{6}\lambda_{S\Phi}\right)T^2 \ .
\end{align}
As above, these can be applied to the pure inert doublet model (without
the singlet) by setting $\lambda_{SH} = \lambda_{S\Phi}= \lambda_S = 0$.


\providecommand{\href}[2]{#2}\begingroup\raggedright\endgroup

\end{document}